\documentclass[12pt, a4paper]{article}
\pdfoutput=1
 \usepackage[font=small,format=plain,labelfont=bf,up,textfont=normal,up,justification=justified,singlelinecheck=false]{caption}

\usepackage[a4paper, left=2cm,right=2cm]{geometry}
\usepackage[colorlinks=true,linkcolor=black,citecolor=teal,urlcolor=MidnightBlue,filecolor=black]{hyperref}
\usepackage{amsfonts}
\usepackage{amsmath}
\usepackage{setspace}
\usepackage[dvipsnames]{xcolor}
\definecolor{SchoolColor}{rgb}{0.6471, 0.1098, 0.1882} 

\usepackage[utf8,applemac]{inputenc}
\usepackage{tensor}
\usepackage{cite}
\usepackage{tikz}
\usepackage{graphicx}
\usepackage{bm} 

\usepackage[utf8,applemac]{inputenc}
\usepackage{tensor}
\usepackage{cite}
\usepackage{tikz}
\usepackage{graphicx,subfigure}
\usepackage[font=small, font+=it, format=hang, margin={1cm, 1cm}]{caption}
\graphicspath{{figure/}}
\usepackage{array}
\usepackage{booktabs} 
\usepackage{multirow}

\usepackage{dcolumn}
\usepackage{bm}

\numberwithin{equation}{section}
\newcommand{\bea}{\begin{eqnarray}}
\newcommand{\eea}{\end{eqnarray}}
\newcommand{\be}{\begin{equation}}
\newcommand{\ee}{\end{equation}}
\def\nn{\nonumber}

\newcommand{\beqs}{\begin{eqnarray}}
\newcommand{\eeqs}{\end{eqnarray}}
\numberwithin{equation}{section}

\begin{document}
\begin{titlepage}

\begin{flushright}\vspace{-3cm}
{\small
\today }\end{flushright}
\vspace{0.5cm}
\begin{center}
	{{ \LARGE{\bf{Area law of 
					connected correlation function \vspace{6pt} in higher dimensional conformal field theory}}}}\vspace{5mm}

	\centerline{\large{\bf{Jiang Long\footnote{e-mail:
					 longjiang@hust.edu.cn}}}}
	\vspace{2mm}
	\normalsize
	\bigskip\medskip

	\textit{School of Physics, Huazhong University of Science and Technology, \\Wuhan, Hubei 430074, China
	}
	
	\vspace{25mm}
	
	\begin{abstract}
		\noindent
		{We present a new area law which is associated with the correlator of OPE blocks in higher dimensional conformal field theories (CFTs).  The area law shows similar behaviour as black hole entropy or geometric entanglement entropy.  It includes a leading term which is proportional to the area of the entanglement surface, and a logarithmic subleading term with degree $q$. We extract the UV cutoff independent coefficients and discuss various properties of the coefficients.}\end{abstract}
	

\end{center}

\end{titlepage}
\tableofcontents

\section{Introduction}
There are diverse area laws in different branches of physics. The prototype is originated from black hole physics where the thermal entropy of a black hole is proportional to the area of its event horizon \cite{Bekenstein:1973ur,Hawking:1974sw}. This unusual property of black hole has stimulated varies modern idea of theoretical physics. 

In the context of quantum field theory (QFT), people have already noticed a similar area law for geometric entanglement entropy \cite{Bombelli:1986rw,Srednicki:1993im,Callan:1994py,Araki:1976zv} several decades ago. One could find the details in the review paper \cite{Eisert:2008ur}. Its connection to gravity has been established by the work of Ryu and Takayanagi \cite{Ryu:2006bv}, in which they proposed that the entanglement entropy of a CFT is equal to the area of a minimal surface in the bulk AdS spacetime. 

In this paper, we present a new area law in general higher dimensional CFTs ($d>2$) following the work \cite{Long:2020njs}. In that work, the author argued that $(m)$-type CCF \cite{Long:2019fay,Long:2019pcv} of OPE block may obey area law from the analytic continuation of $(m-1,1)$-type CCF.  Just like entanglement entropy of continues QFT, it is divergent. The leading term  obeys area law whose coefficient depends on the energy scale. In the sub-leading terms, cutoff independent information can be extracted, usually, this is encoded in a logarithmic divergent term.  However, the logarithmic structure turns out to be much more richer than entanglement entropy.  We summarize the area law and logarithmic behaviour schematically in the following formula 
\be
\langle Q_A[\mathcal{O}_1]\cdots Q_A[\mathcal{O}_m]\rangle_c=\gamma\frac{R^{d-2}}{\epsilon^{d-2}}+\cdots+p_q \log^q\frac{R}{\epsilon}+p_{q-1}\log^{q-1}\frac{R}{\epsilon}+\cdots.\label{log}
\ee 
In this equation, $Q_A[\mathcal{O}]$ is an OPE block associated with a primary operator $\mathcal{O}$. We will review the definition of OPE block in the following section. The subscript $A$ denotes the spacetime region where the OPE block lives in. The quantity $R$ is the  typical size of region A. The small positive parameter $\epsilon$ is a UV cutoff. The constant $\gamma$ is cutoff dependent, therefore it is not physical.  The integer $q$ is the maximal power of the logarithmic terms in the CCF whose coefficient $p_d$ is non-zero. The exact value of $q$ may depend on the  OPE block and the spacetime dimension. According to the value of $q$, we classify the logarithmic behaviour of $(m)$-type CCFs. We will detail its value in the following sections. When the positive value $m\le3$, we find that $q$ may be in the region 
\be
0\le q\le 2.
\ee
The $\cdots$ terms in the formula are the  possible sub-leading terms which are cutoff dependent. Therefore we will not be careful about their exact forms. The physical information is encoded in the coefficient $p_q$.  We establish a UV/IR relation to extract the coefficient $p_q$ based on the analytic continuation of conformal block.

This paper is organised as follows. We begin by introducing OPE block and CCF used in this work in section 2.  In section 3 we will derive the area law and logarithmic behaviour of $(m)$-type CCF. We classify different CCFs according to the maximal power $q$ of the logarithmic term in the CCFs. At the same time, we obtain a UV/IR relation which is useful to extract the cutoff independent coefficient $p_d$. We compute several examples in the following section. In section 5, we discuss an ``inconsistency'' problem and solve it partly.  Section 6 contains some concluding remarks in this work.
\section{Setup}
\subsection{OPE block}
In CFTs, operators are classified into (quasi-)primary operators $\mathcal{O}$ and their descendants $\partial_{\mu}\partial_\nu \cdots \mathcal{O}$. A general primary operator is characterized by two quantum numbers, conformal weight $\Delta$ and spin $J$. Under a global conformal transformation $x\to x'$, a primary operator\footnote{We use scalar field as an example.} transforms as 
\be
\mathcal{O}(x)\to |\frac{\partial x'}{\partial x}|^{-\Delta/d}\mathcal{O}(x).
\ee 
where $|\partial x'/ \partial x|$ is the Jacobian of the conformal transformation of the coordinates, $\Delta$ is the conformal weight of the primary operator and $d$ is the spacetime dimension.  Operator product expansion(OPE) of two separated primary scalar operators $\mathcal{O}_i(x_1)\mathcal{O}_j(x_2)$ is to expand it in a local complete basis around at a suitable point 
\be
\mathcal{O}_i(x_1)\mathcal{O}_j(x_2)=\sum_{k}C_{ijk}|x_{12}|^{\Delta_k-\Delta_i-\Delta_j}(\mathcal{O}_k(x_2)+\cdots),\label{ope}
\ee 
where $\cdots$ are descendants of the primary operator $\mathcal{O}_k$. Its form is fixed by global conformal symmetry, therefore it just contains kinematic information of the CFT. The summation is over all possible parimary operators in the CFT. The constants $C_{ijk}$ are called OPE coefficients which is related to the three point function of the primary operators. They are the only dynamical parameters in the theory.  The constants $\Delta_i, \Delta_j,\Delta_k$ are conformal weights of the corresponding primary operators. The distance of the two points $x_1$ and $x_2$ is denoted as $|x_{12}|$. By collecting all kinematic terms in the summation, we can rewrite the OPE \eqref{ope} as 
\be
\mathcal{O}_i(x_1)\mathcal{O}_j(x_2)=|x_{12}|^{-\Delta_i-\Delta_j}\sum_k C_{ijk}Q^{ij}_k(x_1,x_2).
\ee
The objects $Q^{ij}_k(x_1,x_2)$ are called OPE blocks \cite{Ferrara:1971vh,Ferrara:1972cq,Czech:2016xec}. They are non-local operators in the CFT and  depend on the position of external operators $x_1$ and  $x_2$. The upper index $i$ and $j$ show that it also depends on the quantum number of the external operators $\mathcal{O}_i$ and $\mathcal{O}_j$. It is easy to see that the OPE block has dimension zero. Under a global conformal transformation $x\to x'$, an OPE block $Q^{ij}_k(x_1,x_2)$ will transform as 
\be
Q^{ij}_k(x_1,x_2)\to f(x_1',x_2')Q^{ij}_k(x_1',x_2').
\ee 
The explicit form of $f(x_1',x_2')$ is not important in this work. When the two external operators are the same, we have $f(x_1',x_2')=1$ and the OPE block will be invariant under global conformal transformations. One can also show that the OPE block is independent of the external operator in this special case. We will relabel such kind of OPE block as 
\be
Q_A[\mathcal{O}_k]=Q_k^{ii}(x_1,x_2).\label{opeii}
\ee 
The subscript $A$ denotes the region determined by the two points $x_1$ and $x_2$ where the two external operators insert into.  The operator in the square bracket reflects the fact that the OPE block is generated by the primary operator $\mathcal{O}_k$. We omit the information of $i$ since this OPE block is insensitive to the external operators.  We will classify the primary operators $\mathcal{O}_k$ into conserved currents $\mathcal{J}$ and non-conserved operators $\mathcal{O}$. A general primary operator obeys the following unitary bound \cite{Minwalla:1997ka}\footnote{The operators in this work are symmetric and traceless. We will also not discuss fermion operators.}
\bea
\left\{\begin{array}{l}
\Delta\ge J+d-2,\quad J\ge 1,\nn\\
\Delta\ge \frac{d-2}{2},\quad J=0.\end{array}
\right.
\eea
A conserved current $\mathcal{J}$ with spin $J (J\ge 1)$ will satisfy $\Delta=J+d-2$.  All other primary operators are non-conserved operators. 
Correspondingly, the OPE block \eqref{opeii} generated by a conserved current $\mathcal{J}$  will be called a type-J OPE block.  On the other hand, the OPE block \eqref{opeii} generated by a non-conserved operator $\mathcal{O}$  will be called a type-O OPE block.  

When the two external operators are time-like separated, the region $A$ is a causal diamond. The two operators are at the sharp corner of the diamond $A$.  We can use conformal transformations to fix 
\be
x_1=(1, \vec{x}_A),\quad x_2=(-1,\vec{x}_A), \label{x1x2}
\ee 
then the causal diamond $A$ intersects $t=0$ slice with a unit  ball $\Sigma_A$
\be
\Sigma_A=\{(0,\vec{x})|(\vec{x}-\vec{x}_A)^2\le 1\}.\label{sigma}
\ee
The center of the  ball is $\vec{x}_A$. The boundary of $\Sigma_A$ is a unit sphere $S$.  In the context of geometric entanglement entropy, the surface $S$ is an entanglement surface which separates the ball $\Sigma_A$ and its complement. The leading term of the entanglement entropy is proportional to the area of the surface $S$ in general higher dimensions ($d>2$). There is a conformal Killing vector $K$ which preserves the diamond $A$
\be
K^{\mu}=\frac{1}{2}(1-(\vec{x}-\vec{x}_A)^2-t^2,-2t \vec{x})
\ee 
and it is null on the boundary of the diamond $A$. It generates a modular flow of the diamond $A$.  Any type-O OPE block corresponds to the point pair \eqref{x1x2} or the unit ball $\Sigma_A$ \eqref{sigma} is \cite{deBoer:2016pqk}
\be
Q_A[\mathcal{O}_{\mu_1\cdots\mu_J}]=c_{\mathcal{O}_{\mu_1\cdots\mu_J}}\int_{A} d^dx K^{\mu_1}\cdots K^{\mu_J}|K|^{\Delta-d-J} \mathcal{O}_{\mu_1\cdots \mu_J},\label{typeO}
\ee 
where the primary operator $\mathcal{O}_{\mu_1\cdots\mu_J}$ is non-conserved 
\be
\partial^{\mu_1}\mathcal{O}_{\mu_1\cdots\mu_J}\not=0.
\ee 
It has dimension $\Delta$ and spin $J$.  When the operator is a conserved current 
\be
\partial^{\mu_1}\mathcal{J}_{\mu_1\cdots\mu_J}=0,\label{cons}
\ee
the corresponding type-J OPE block is
\be
Q_A[\mathcal{J}_{\mu_1\cdots\mu_J}]=c_{\mathcal{J}_{\mu_1\cdots\mu_J}}\int_{\Sigma_A}d^{d-1}\vec{x} (K^0)^{J-1}\mathcal{J}_{0\cdots0}. \label{typeJ}
\ee 
It can be obtained from \eqref{typeO} by using conservation law \eqref{cons} and reducing it to a lower $d-1$ dimensional integral. The coefficient $c_{\mathcal{J}_{\mu_1\cdots\mu_J}}$ is also redefined at the same time. In \eqref{typeO} and \eqref{typeJ}, the coefficients $c_{\mathcal{O}_{\mu_1\cdots\mu_J}}$ and $c_{\mathcal{J}_{\mu_1\cdots\mu_J}}$ are free parameters, we set them to be 1. 

A very special type-J OPE block is the modular Hamiltonian \cite{Haag:1992,Casini:2011kv} of the region $\Sigma_A$, 
\be
H_A=2\pi \int_{\Sigma_A}d^{d-1}\vec{x} K^0 T_{00}=2\pi \int_{\Sigma_A}d^{d-1}\vec{x} \frac{1-(\vec{x}-\vec{x}_A)^2}{2} T_{00}(0,\vec{x}).
\ee
The modular Hamiltonian is the logarithm of the reduced density matrix $\rho_A$
\be
H_A=-\log \rho_A.\label{mod}
\ee 
It plays a central role in the context of 
entanglement entropy, 
\be
S_A=-\text{tr}_{A}\rho_A \log\rho_A=\text{tr}_{A} e^{-H_A}H_A.
\ee 
More generally, R\'enyi entanglement entropy
\be
S_A^{(n)}=\frac{1}{1-n}\log\text{tr}_{A}\rho_A^n
\ee 
has been shown to satisfy an area law generally 
\be
S_A^{(n)}=\gamma \frac{\mathcal{A}}{\epsilon^{d-2}}+\cdots,
\ee 
where $\mathcal{A}$ is the area of the entanglement surface $S$  and $\epsilon$ is a UV cutoff.  The constant $\gamma$ is cutoff dependent.  The subleading terms $\cdots$ contain a logarithmic term  in even dimensions
\be
S_A^{(n)}=\gamma  \frac{\mathcal{A}}{\epsilon^{d-2}}+\cdots+p_1(n)\log\frac{R}{\epsilon}+\cdots,\label{nrenyi}
\ee
where we have inserted back the radius $R$. The area $\mathcal{A}$ is related to the radius $R$ through the power law 
\be
\mathcal{A}\sim R^{d-2}.
\ee
The coefficient $p_1(n)$ encodes useful information of the CFT.  It is easy to show that the CCF of the modular Hamiltonian $H_A$ satisfies a similar area law in even dimensions,
\be
\langle H_A^m\rangle_c=\tilde{\gamma}\frac{\mathcal{A}}{\epsilon^{d-2}}+\cdots+\tilde{p}_1^{(m)} \log\frac{R}{\epsilon}+\cdots,\quad m\ge 1.\label{HAm}
\ee 
The coefficient $\tilde{p}_1^{(m)}$ is determined from $p_1(n)$ by
\be
\tilde{p}_1^{(m)}=(-1)^m\partial_n^{m}(1-n)p_1(n) |_{n\to 1}.
\ee
We will introduce the definition of the CCF in the following subsection. 

\subsection{Deformed reduced density matrix and connected correlation 
function}
Reduced density matrix of a subregion $A$ is obtained  by tracing out the degree of freedom in its complement 
\be
\rho_A=\text{tr}_{\bar{A}}\rho
\ee 
where $\rho$ is the density matrix of the system. It can also be written as an exponential operator formally \eqref{mod}
\be
\rho_A=e^{-H_A}.
\ee
For a causal diamond $A$, $H_A$ is a type-J OPE block. Therefore it is natural to define a deformed reduced density matrix \cite{Long:2019pcv} by replacing the modular Hamiltonian with a general OPE block $Q_A$ 
\be
\rho_A=e^{-\mu Q_A},\label{drd}
\ee
where we still use $\rho_A$ to label deformed reduced density matrix. The constant $\mu$ is an independent constant. In the ``first law of thermodynamics''\cite{Long:2020njs} associated with the deformed reduced density matrix, it may be regarded as a chemical potential which is dual to the OPE block $Q_A$. The OPE block $Q_A$  can also be a linear superposition of multiple OPE blocks.  We don't restrict the OPE block in \eqref{drd} to be type-J. A subtle problem is that the spectrum of $Q_A$ is not always non-negative, therefore the deformed reduced density matrix  may not be well defined in general. However, as we will show below, it is  still a useful formal tool to generate CCFs.

We define a formal generator of the (m)-type CCF through the logarithm of the vacuum expectation value of the deformed reduced density matrix, 
\be
T_A(\mu)=\log \langle e^{-\mu Q_A}\rangle.
\ee 
Then the so-called (m)-type CCF of the OPE block $Q_A[\mathcal{O}]$ is defined as 
\be
\langle Q_A[\mathcal{O}]^m\rangle_c=(-1)^m \frac{\partial^mT_A(\mu)}{\partial \mu^m}\big{|}_{\mu\to0}
\ee 
The first few orders are 
\bea
\langle Q_A[\mathcal{O}]^2\rangle_c&=&\langle Q_A[\mathcal{O}]^2\rangle-\langle Q_A[\mathcal{O}]\rangle^2,\nn\\
\langle Q_A[\mathcal{O}]^3\rangle_c&=&\langle Q_A[\mathcal{O}]^3\rangle-3\langle Q_A[\mathcal{O}]^2\rangle\langle Q_A[\mathcal{O}]\rangle+2\langle Q_A[\mathcal{O}]\rangle^3.
\eea 
When there are multiple space-like separated regions, one can define a general $Y$-type CCF with the Young diagram 
\be
Y=(m_1,m_2,\cdots),\quad m_1\ge m_2\ge\cdots\ge1.
\ee 
The OPE block generated from the operator $\mathcal{O}$ is an eigenvector of the Casimir operator of the conformal group with the eigenvalue $C=\Delta(\Delta-d)-J(J+d-2)$. Combining with the boundary behaviour when $x_1\to x_2$ for the OPE block,  any $(m,1)$-type CCF will be proportional to a conformal block 
\be
\langle Q_A[\mathcal{O}]^m Q_B[\mathcal{O}]\rangle_c=D[\mathcal{O}]G_{\Delta,J}(z),\label{mm1}
\ee 
where $B$ is another causal diamond, $z$ denotes the cross ratios corresponding to the two diamonds $A$ and $B$. 
The OPE blocks can be different in \eqref{mm1}, we write the general result as 
\be
\langle Q_A[\mathcal{O}_1]\cdots Q_A[\mathcal{O}_m] Q_B[\mathcal{O}]\rangle_c=D[\mathcal{O}_1,\mathcal{O}_2,\cdots,\mathcal{O}_m,\mathcal{O}]G_{\Delta,J}(z).\label{12m}
\ee 
The coefficient $D$ characterizes the large distance behaviour of $(m,1)$-type CCF. The references to discuss conformal block are \cite{Dolan:2000ut,Dolan:2003hv}. In this work, we just need the diagonal limit of the conformal block \cite{Hogervorst:2013kva}.
\section{Area law and logarithmic behaviour}
Motivated by the area law of R\'enyi entanglement entropy \eqref{nrenyi}, or equivalently, the area law of the $(m)$-type CCF of the modular Hamiltonian \eqref{HAm}, we are interested in the divergent behaviour of  the $(m)$-type CCF of OPE blocks
\be
\langle Q_A[\mathcal{O}_1]\cdots Q_A[\mathcal{O}_m]\rangle_c.\label{QAm}
\ee 
When the OPE block is the modular Hamiltonian, we should reproduce the area law of modular Hamiltonian \eqref{HAm}. Therefore it is natural to conjecture that \eqref{QAm} also obeys an area law for general OPE blocks. In the subleading terms, one may also read out cutoff independent information. It turns out that the structure is much more richer, 
\be
\langle Q_A[\mathcal{O}_1]\cdots Q_A[\mathcal{O}_m]\rangle_c=\gamma\frac{R^{d-2}}{\epsilon^{d-2}}+\cdots+p_q \log^q\frac{R}{\epsilon}+p_{q-1}\log^{q-1}\frac{R}{\epsilon}+\cdots.\label{logq}
\ee 
As discussed in the introduction, the maximal power of $\log\frac{R}{\epsilon}$ is $q$. We will call $q$ the degree of the $(m)$-type CCF \eqref{QAm}. For example, the degree $q$ is one for the CCF of the modular Hamiltonian \eqref{HAm} or \eqref{nrenyi} in even dimensions. In this paper, we will restrict the integer $m\le 3$, then the degree may satisfy $0\le q\le 2$. More explicitly, 
\bea
q=\left\{\begin{array}{l}1,2,\quad d=\text{even}\\
	0,1,\quad d=\text{odd}\end{array}\right.\label{qbe}
\eea
We will use the degree $q$ to distinguish CCFs \eqref{QAm}. In the following, we will discuss the logarithmic behaviour in detail.

In even dimensions, as \eqref{qbe},  we could distinguish two classes according to the logarithmic behaviour in the subleading terms.
\begin{enumerate}
\item Class I. The degree of the $(m)$-type CCF is 1. We can write \eqref{logq} more explicitly as 
\be
\langle Q_A[\mathcal{O}_1]\cdots Q_A[\mathcal{O}_m]\rangle_c=\gamma[\mathcal{O}_1,\cdots,\mathcal{O}_m] \frac{R^{d-2}}{\epsilon^{d-2}}+\cdots+p^e_1[\mathcal{O}_1,\cdots,\mathcal{O}_m] \log\frac{R}{\epsilon}+\cdots,\label{Jme}
\ee 
where we detail the dependence of the primary operator $\mathcal{O}_i $ for the coefficients $\gamma$ and $p_1$.  The upper index $e$ in $p_1$ indicates that the spacetime dimension is even. The well known example is the CCF of the modular Hamiltonians \eqref{HAm}, or equivalently \eqref{nrenyi}. For simplicity, we set the spacetime dimension $d=4$. There are many discussions on the structure \eqref{nrenyi} or \eqref{HAm}.  We will argue the structure \eqref{HAm} in the following way. We'd like to make use of the conclusion \eqref{12m} by moving one OPE block to a separated region $B$, then the left hand side of \eqref{Jme} becomes a $(m-1,1)$-type CCF 
\be
\langle H_A^{m-1}H_B\rangle_c=D[T_{\mu_1\nu_1},\cdots,T_{\mu_m\nu_m}]G_{4,2}(z).\label{HAm1HB}
\ee
We can choose the region $B$ as the causal diamond of a unit ball $\Sigma_B$ whose radius is $R'$.
\be
\Sigma_B=\{(0,\vec{x})|\vec{x}^2\le R'^2\}.
\ee 
The center of the ball is origin. 
Therefore the unique cross ratio of $\Sigma_A$ and $\Sigma_B$ is \footnote{Usually, there are two cross ratios for two balls. However, $\Sigma_A$ and $\Sigma_B$ are located at the same time $t=0$ which reduce the number of independent of cross ratio to one. }
\be
z=\frac{4R'}{x_A^2-(1-R')^2}.
\ee 
The conformal block $G_{4,2}(z)$ is well defined for $0<z<1$, which is exactly the case that $A$ and $B$ are space-like separated.  Now we move the diamond $B$ to $A$, then the $(m-1,1)$-type CCF becomes an $(m)$-type CCF. Roughly speaking 
\be
\langle H_A^m\rangle_c=\text{lim}_{B\to A} \langle H_A^{m-1}H_B\rangle_c.
\ee 
The limit $B\to A$ is subtle, we first move $x_A\to0$ and then take the limit $R'\to1$, 
\be
r=0,\quad R'=1-\epsilon, \quad \epsilon\to 0.
\ee 
The cross ratio $z$ approaches $-\infty$ by 
\be
z=-\frac{4(1-\epsilon)}{\epsilon^2},\quad \epsilon\to 0.\label{z-inf}
\ee 
In this limit, the conformal block $G_{4,2}(z)$ becomes divergent 
\be
G_{4,2}(z)\to \tilde{\gamma} \frac{R^2}{\epsilon^2}+\cdots-120\log\frac{R}{\epsilon}+\cdots.
\ee 
We  have inserted back the radius $R$ in the expression. The leading term is proportional to area of the surface $S$. As $B$ approaches $A$,  the 
$(m-1,1)$-type CCF becomes a $(m)$-type CCF
\be
\langle H_A^m\rangle_c=\gamma\frac{R^2}{\epsilon^2}+\cdots+p_1^e[T_{\mu_1\nu_1},\cdots,T_{\mu_m\nu_m}]\log\frac{R}{\epsilon}+\cdots \label{HAm2}
\ee 
with 
\be
p_1^e[T_{\mu_1\nu_1},\cdots,T_{\mu_m\nu_m}]=-120 D[T_{\mu_1\nu_1},\cdots,T_{\mu_m\nu_m}].\label{p1es2}
\ee 
If the coefficient $D$ is finite in \eqref{p1es2}, then \eqref{HAm2} is exactly the same as \eqref{HAm}. 
The equation \eqref{p1es2} is a typical UV/IR relation for modular Hamiltonian in the sense of \cite{Long:2020njs}. The left hand side is the cutoff independent coefficient as $B$ and $A$ coincides (UV) while $D$ characterizes the leading order behaviour of CCF when two regions are far away to each other (IR). The constant $-120$  is  from the conformal block associated with the stress tensor in four dimensions. Therefore it is a kinematic term which is totally fixed by conformal symmetry.  Note the constant $\gamma$ is cutoff dependent, therefore it may depend on the energy scale we choose. 

The discussion on modular Hamiltonian may extend to other OPE blocks. Interestingly, we find that a conformal block $G_{\Delta,J}(z)$ in even dimensions has either degree $q=1$ or $q=2$
\bea
G_{\Delta,J}(z)\sim \left\{\begin{array}{l}\tilde{\gamma}\frac{R^{d-2}}{\epsilon^{d-2}}+\cdots +E[\Delta,J]\log \frac{R}{\epsilon}+\cdots,\quad \Delta=J+d-2,\vspace{4pt}\\
	\tilde{\gamma}\frac{R^{d-2}}{\epsilon^{d-2}}+\cdots +E[\Delta,J]\log^2 \frac{R}{\epsilon}+\cdots,\Delta>J+d-2,\vspace{4pt}\\
	\end{array}\right.
\eea  
where the constant $E[\Delta,J]$ is determined by quantum numbers of the primary operator. When all the primary operators are conserved currents, $\Delta=J+d-2$, we conclude that the $(m)$-type CCF of type-J OPE blocks may has degree $q=1$ with 
\be
p_1^e[\mathcal{O}_1,\cdots,\mathcal{O}_m]=E[\mathcal{O}_m]D[\mathcal{O}_1,\cdots,\mathcal{O}_m], \label{UVIR1}
\ee
where we have replaced the quantum numbers in $E$ function by the primary operator. 
Some remarks are shown as follows. 
\begin{enumerate}
\item Cyclic identity. For a general $(m)$-type CCF of the type-J OPE block \eqref{Jme}, we have different ways to uplift $(m)$-type to $(m-1,1)$ type. However, the cutoff independent coefficient should be equal since they lead to the same CCF. For example, $m=3$, the coefficients $p_1^e$ should satisfy the following cyclic identity
\be
p_1^e[\mathcal{O}_2,\mathcal{O}_3,\mathcal{O}_1]=p_1^e[\mathcal{O}_3,\mathcal{O}_1,\mathcal{O}_2]=p_1^e[\mathcal{O}_1,\mathcal{O}_2,\mathcal{O}_3].\label{constr}
\ee 
\item 
The function $E[\mathcal{O}]$ can be read out from the conformal block $G_{\Delta,J}$ corresponding to the primary operator $\mathcal{O}$. For conserved currents, we find
\bea
E[\mathcal{\mathcal{O}}]=\left\{\begin{array}{l}
12,\quad \quad \ \Delta=3, J=1,\vspace{4pt}\\
-120,\quad \Delta=4,J=2,\vspace{4pt}\\
840,\quad \ \ \Delta=5,J=3,\vspace{4pt}\\
\cdots
\end{array}\right.\label{EJ}
\eea 
\item The constant $\gamma$ also depends on the way to uplift $(m)$-type CCF to $(m-1,1)$ type. Since it is cutoff dependent, we don't expect they are equal to each other, 
\be
\gamma[\mathcal{O}_2,\cdots,\mathcal{O}_m,\mathcal{O}_1]\not=\gamma[\mathcal{O}_1,\mathcal{O}_3,\cdots,\mathcal{O}_m,\mathcal{O}_2]\not=\cdots\not=\gamma[\mathcal{O}_1,\cdots,\mathcal{O}_{m-1},\mathcal{O}_m].
\ee 
\end{enumerate}
\item Class II. For this class, the degree $q=2$, 
\be
\langle Q_A[\mathcal{O}_1]\cdots Q_A[\mathcal{O}_m]\rangle_c=\gamma \frac{R^{d-2}}{\epsilon^{d-2}}+\cdots+p^e_2[\mathcal{O}_1,\cdots,\mathcal{O}_m]\log^2\frac{R}{\epsilon}+p^e_1[\mathcal{O}_1,\cdots,\mathcal{O}_m] \log\frac{R}{\epsilon}+\cdots.
\ee
Therefore the coefficient $p_2^e$ is cutoff independent while $p_1^e$ is not. As Class I, 
  we can read UV/IR relation 
\be
p_2^e[\mathcal{O}_1,\cdots,\mathcal{O}_{m-1},\mathcal{O}_m]=E[\mathcal{O}_m]D[\mathcal{O}_1,\cdots,\mathcal{O}_{m-1},\mathcal{O}_m].\label{p2emO}
\ee The coefficient $p_2^e$ should also satisfy a cyclic property as \eqref{constr},
\be
p_2^e[\mathcal{O}_2,\mathcal{O}_3,\mathcal{O}_1]=p_2^e[\mathcal{O}_3,\mathcal{O}_1,\mathcal{O}_2]=p_2^e[\mathcal{O}_1,\mathcal{O}_{2},\mathcal{O}_3].\label{p2eeq}
\ee 
 We read $E[\mathcal{O}]$ from the conformal block $G_{\Delta,J}(z)$ for non-conserved operators, several examples are shown below 
\bea
E[\mathcal{O}]=\left\{\begin{array}{l}
-\frac{2^{2\Delta-1}\Gamma(\frac{\Delta-1}{2})\Gamma(\frac{\Delta+1}{2})}{\pi \Gamma(\frac{\Delta-2}{2})^2},\quad \quad \hspace{5pt}\Delta>1, \quad J=0,\\\vspace{4pt}
\frac{2^{2\Delta-1}\Gamma(\frac{\Delta}{2})\Gamma(\frac{\Delta+2}{2})}{\pi\Gamma(\frac{\Delta-3}{2})\Gamma(\frac{\Delta+1}{2})},\quad \quad\quad\hspace{14pt} \Delta>3,\quad J=1,\\\vspace{4pt}
-\frac{4^{\Delta-1}(\Delta-2)\Gamma(\frac{\Delta-3}{2})\Gamma(\frac{\Delta+3}{2})}{\pi\Gamma(\frac{\Delta-4}{2})\Gamma(\frac{\Delta+2}{2})},\quad \Delta>4,\quad J=2,\vspace{4pt} \\
\cdots
\end{array}\right.
\eea 
There are some constraints on the conformal weight. For scalar primary operator, the unitary bound in four dimensions will  constrain $\Delta\ge 1$. We notice that the function $E[\mathcal{O}]$ becomes divergent when $\Delta=1$. On the other hand, when $\Delta=2$, the function $E[\mathcal{O}]$ is zero. Therefore we should be careful with the two special points.  Since the physical coefficient is the product of $E$ and $D$, see \eqref{p2emO}, we cannot make the conclusion that $p_2^e$ is divergent for $\Delta=1$ and zero for $\Delta=2$ since it also depends on the behaviour of the function $D$  near the two special points. When the non-conserved operators have spin $J\ge 1$, the unitary condition constrains 
\be
\Delta>J+2
\ee 
for CFT$_4$. This is the inequality at the second and third line of $E[\mathcal{O}]$. We also note that as $\Delta\to J+2$, $E[\mathcal{O}]$ actually approaches zero. If the function $D$ is finite in this limit,  \eqref{p2emO} implies that $p_2^e$ is zero for $\Delta=J+2$. Then $p_1^e$ becomes cutoff independent,  which is consistent with the conclusion in Class I.  
\end{enumerate}

In odd dimensions, the logarithmic behaviour is a bit different, however, we could still distinguish two classes according to the degree $q$. It turns out that the maximal degree $q$ is 1 in odd dimensions.  We discuss them briefly in the following as it is parallel to even dimensions. 
\begin{enumerate}
\item Class O. In this class, the degree $q=0$, 
\be
\langle Q_A[\mathcal{O}_1]\cdots Q_A[\mathcal{O}_m]\rangle_c=\gamma \frac{R^{d-2}}{\epsilon^{d-2}}+\cdots+p^o_0[\mathcal{O}_1,\cdots,\mathcal{O}_m] .
\ee 
There is no logarithmic divergence in this case. The upper index in $p^o_0$  denotes that the spacetime dimension is odd. 
\item Class I'. In this class, the degree $q=1$, 
\be
\langle Q_A[\mathcal{O}_1]\cdots Q_A[\mathcal{O}_m]\rangle_c=\gamma \frac{R^{d-2}}{\epsilon^{d-2}}+\cdots+p^o_1[\mathcal{O}_1,\cdots,\mathcal{O}_m] \log\frac{R}{\epsilon}+\cdots.
\ee
We can also find the corresponding UV/IR relations. For example, in three dimensions,  the function $E[\mathcal{O}]$  is 
\bea
E[\mathcal{O}]=\left\{\begin{array}{l}
-\frac{2^{2\Delta-1}(\Delta-1)\Gamma(\Delta-\frac{1}{2})}{\sqrt{\pi}\Gamma(\Delta-1)},\quad\hspace{5pt} \Delta>\frac{1}{2}, \quad J=0.\\
\vspace{4pt}
\frac{2^{\Delta+1}\Delta\Gamma(\Delta-\frac{1}{2})}{\Gamma(\frac{\Delta-2}{2})\Gamma(\frac{\Delta+1}{2})},\quad\hspace{34pt} \Delta>2,\quad J=1,\\\vspace{4pt}
-\frac{2^{2\Delta-1}(\Delta^2-1)\Gamma(\Delta-\frac{1}{2})}{\sqrt{\pi}(\Delta-2)^2\Delta\Gamma(\Delta-3)},\quad \Delta>3,\quad J=2,\vspace{4pt}\\
\cdots
\end{array}\right.
\eea 
\end{enumerate}

\section{Examples}
In this section, we will use several examples to check the results in the previous section. We will set spacetime dimension $d=4$ from now on. 
\subsection{Class I}
Type-J OPE block is 
\bea
Q_{A}[\mathcal{J}_{\mu_1\cdots\mu_J}]=\int_{\Sigma_A} d^{3}\vec{x} (K^0)^{J-1}\mathcal{J}_{0\cdots0}=\frac{1}{2^{J-1}}\int_{\Sigma_A}d^{d-1}\vec{x} (1-(\vec{x}-\vec{x}_A)^2)^{J-1}\mathcal{J}_{0\cdots0}.
\eea 
\subsubsection{$(2)$-type}
We will consider conserved currents with lower spin $J\le 2$. 
\begin{enumerate}
\item 
Spin 1 current. We will use two methods to compute the CCF
\bea
\langle Q_A[\mathcal{J}_{\mu}]^2\rangle_c=\ :\int_{\Sigma_A} d^{3}\vec{x} \int_{\Sigma_A}d^{3}\vec{x}' \langle \mathcal{J}_0(\vec{x})\mathcal{J}_0(\vec{x'})\rangle_c:.\label{spin1}
\eea 
The symbol $:\ :$ means that one should remove the divergence from the two operators attach to each other \cite{Long:2019fay}. This requires a way of regularization. In the following, we will omit the symbol $:\ :$. 
\begin{enumerate}
\item We transform the coordinates to spherical coordinates 
\be
\vec{x}=r \vec{\omega},\quad \vec{\omega}^2=1, 
\ee 
then
\bea
&&\langle Q_A[\mathcal{J}_{\mu}]^2\rangle_c\nn\\&=&\int_{\Sigma_A} r^{2}dr d\vec{\omega} \int_{\Sigma_A}r'^2 dr' d\vec{\omega'}\frac{C_{J}I_{00}(x-x')}{|\vec{x}-\vec{x}'|^6}\nn\\&=&-C_J \int_{\Sigma_A}r^{2}dr d\vec{\omega} \int_{\Sigma_A}r'^2 dr' d\vec{\omega'}\frac{1}{(r^2+r'^2-2r r' \vec{\omega}\cdot\vec{\omega}')^3}\nn\\&=&-C_J S_2 S_1 \int_0^1 r^2 dr \int_0^1 r'^2 dr' \int_0^{\pi}\sin\theta d\theta \frac{1}{(r^2+r'^2-2r r' \cos\theta)^3}\nn\\&=&-C_J (4\pi)\times (2\pi)\int_0^1 dr \int_0^1 dr' \frac{2r^2r'^2(r^2+r'^2)}{(r-r')^4(r+r')^4}\nn\\&=&-8\pi^2C_J \int_0^{1-\epsilon} dr \frac{2r^2}{3(r^2-1)^3}\nn\\
&=&\frac{\pi^2}{3}C_J (\frac{R^2}{\epsilon^2}-\frac{R}{\epsilon}-\log\frac{R}{\epsilon}+\cdots).\label{QAJ}
\eea 
At the first step, we make use of the two point function of the spin 1 current 
\be
\langle \mathcal{J}_{\mu}(x)\mathcal{J}_{\nu}(x')\rangle=\frac{C_J I_{\mu\nu}(x-x')}{|x-x'|^{2\Delta}},
\ee
where the symmetric tensor is
\be
I_{\mu\nu}(x)=\eta_{\mu\nu}-2n_{\mu}(x)n_{\nu}(x),\quad n_{\mu}=\frac{x_{\mu}}{|x|}.
\ee 
The constant $C_{J}$ defines the normalization of the current $\mathcal{J}_{\mu}$.
At the time slice $t=0$, we have $n_0=0$ and $I_{00}=\eta_{00}$. At the third line, we define the angle $\theta$ between the two vectors $\vec{\omega}$ and $\vec{\omega}'$, 
\be
\vec{\omega}\cdot \vec{\omega}'=\cos\theta.
\ee 
The factor $S_n$ is the area of the unit $n$-sphere  $S^n$,\be
S_n=\frac{2\pi^{\frac{n+1}{2}}}{\Gamma(\frac{n+1}{2})}.\ee The integrand at the fourth line has poles at
\be
r=r'.
\ee 
According to the regularization method in \cite{Long:2019fay}, we can just ignore those poles. These poles are from the two currents $\mathcal{J}_{\mu}(x)$ and $\mathcal{J}_{\nu}(x')$ attach to each other. We expect they can be removed\footnote{In Appendix \ref{sing}, we study carefully the pole structure around the point $r=r'$ and find that they have no contribution to the logarithmic divergence. Therefore they don't affect the cutoff independent coefficient.}. At the fifth line, the integrand is also divergent for $r\to 1$. Therefore we insert a small positive $\epsilon$ into the upper bound of the integration. The small parameter $\epsilon$ characterizes the distance to the entanglement surface, therefore it is a UV cutoff. 
At the last step, we insert back the radius $R=1$ to balance the dimension. The term in $\cdots$ is an unimportant constant.  Now we can extract the cutoff independent coefficient 
\be
p_1^e[\mathcal{J}_{\mu},\mathcal{J}_{\nu}]=-\frac{\pi^2}{3}C_J.\label{p1eJ1}
\ee
\item Now we can also compute the same CCF \eqref{spin1} by uplifting the $(2)$-type CCF to $(1,1)$-type, namely 
\be
\langle Q_A[\mathcal{J}_{\mu}]^2\rangle_c\stackrel{uplift}{\longrightarrow}\langle Q_A[\mathcal{J}_{\mu}]Q_B[\mathcal{J}_{\nu}]\rangle_c
\ee 
The $(1,1)$-type CCF is easy to compute as we just need to fix the leading order coefficient $D[\mathcal{J}_{\mu},\mathcal{J}_{\nu}]$ when $A$ and $B$ are far apart. 
\bea
&&\langle Q_A[\mathcal{J}_{\mu}]Q_B[\mathcal{J}_{\nu}]\rangle_c\nn\\&=&\int_{\Sigma_A} r^{2}dr d\vec{\omega} \int_{\Sigma_B}r'^2 dr' d\vec{\omega'}\frac{C_{J}I_{00}(x-x')}{|\vec{x}+\vec{x}_A-\vec{x}'|^6}\nn\\&\approx&-C_J \int_{\Sigma_A} r^{2}dr d\vec{\omega} \int_{\Sigma_B}r'^2 dr' d\vec{\omega'}\frac{1}{x_A^6}\nn\\&=&-C_J (\frac{4\pi}{3})^2\times \frac{1}{2^6}z^3\nn\\
&=&-\frac{\pi^2}{36}C_Jz^3.
\eea 
At the first step, we insert back the center of $\Sigma_A$. The center of $\Sigma_B$ is assumed to be 0. At the second step, we use the assumption that $A$ and $B$ are far away to each other, $x_A\to\infty$.  At the third step, we rewrite $x_A$ in terms of the cross ratio 
\be
z=\frac{4}{x_A^2}. 
\ee
We read out the value 
\be
D[\mathcal{J}_{\mu},\mathcal{J}_{\nu}]=-\frac{\pi^2}{36}C_J.
\ee
Then we use the UV/IR relation \eqref{UVIR1} and the function $E[\mathcal{J}]=12$ for spin 1 current to obtain
\be
p_1^e[\mathcal{J}_{\mu},\mathcal{J}_{\nu}]=-\frac{\pi^2}{36}C_J\times 12=-\frac{\pi^2}{3}C_J.\label{p1eJ2}
\ee 
\end{enumerate}
As we expect, the coefficients \eqref{p1eJ1} and \eqref{p1eJ2} are the same. 
It is also easy to check that the coefficient $\gamma$ are not the same for the two methods.  Since $\gamma$ has no cutoff independent meaning, it depends on the regularization. One can redefine the cutoff such that they are the same.

\item Spin 2 current. As spin 1 current, we use two ways to regularize the integral. 
\begin{enumerate}
\item The first method is to regularize the integral directly, we need the two point function for spin 2 current 
\be
\langle T_{\mu\nu}(x)T_{\rho\sigma}(x')\rangle=C_T \frac{I_{\mu\nu,\rho\sigma}(x-x')}{|x-x'|^{2\Delta}},
\ee 
where 
\be
I_{\mu\nu,\rho\sigma}(x)=\frac{1}{2}(I_{\mu\rho}(x)I_{\nu\sigma}(x)+I_{\mu\sigma}(x)I_{\nu\rho}(x))-\frac{1}{4}\eta_{\mu\nu}\eta_{\rho\sigma}.
\ee 
At the time slice $t=0$, we find 
\be
I_{00,00}=\frac{3}{4}.
\ee 
Then
\bea
&&\langle Q_A[T_{\mu\nu}]^2\rangle_c\nn\\&=&\frac{1}{4}\times \frac{3}{4}C_T\int_{\Sigma_A}r^2 dr d\vec{\omega} \int_{\Sigma_A}r'^2 dr' d\vec{\omega}'\frac{(1-r^2)(1-r'^2)}{(r^2+r'^2-2r r' \vec{\omega}\cdot\vec{\omega}')^4}\nn\\
&=&\frac{3}{16}S_2 S_1 C_T \int_0^1 dr \int_0^1 dr'\frac{2r^2(1-r^2)r'^2(1-r'^2)(r^2+3r'^2)(r'^2+3r^2)}{3(r^2-r'^2)^6}\nn\\
&=&\frac{\pi^2}{40}C_T (\frac{R^2}{\epsilon^2}-\frac{R}{\epsilon}-\log\frac{R}{\epsilon}+\cdots).\label{QAT}
\eea
We read 
\be
p_1^e[T_{\mu\nu},T_{\rho\sigma}]=-\frac{\pi^2}{40}C_T .\label{p1eT1}
\ee 
\item  We can also compute $(1,1)$-type CCF firstly, 
\bea
&&\langle Q_A[T_{\mu\nu}Q_B[T_{\rho\sigma}]\rangle_c\nn\\&\approx&\frac{1}{4}\times \frac{3}{4}C_T\int_{\Sigma_A}r^2 dr d\vec{\omega} \int_{\Sigma_A}r'^2 dr' d\vec{\omega}' (1-r^2)(1-r'^2)\frac{1}{x_A^8}\nn\\
&=&\frac{\pi^2}{4800}C_T z^4.
\eea
Therefore we get 
\be
D[T_{\mu\nu},T_{\rho\sigma}]=\frac{\pi^2}{4800}C_T.
\ee 
Combining with $E[T_{\mu\nu}]=-120$ for the stress tensor and the UV/IR relation, 
\be
p_1^e[T_{\mu\nu},T_{\rho\sigma}]=\frac{\pi^2}{4800}C_T\times (-120)=-\frac{\pi^2}{40}C_T. \label{p1eT2}
\ee 
\end{enumerate}
Again, we find the cutoff independent coefficients \eqref{p1eT1} and \eqref{p1eT2} are equal. We note that
$\langle Q_A[T_{\mu\nu}]^2\rangle_c$ is related to the universal property of R\'enyi entanglement entropy by \cite{Perlmutter:2013gua}. Transforming to the notation of that paper, we have 
\bea
&&\langle Q_A[T_{\mu\nu}]^2\rangle_c=\langle H_\tau^2\rangle=-\frac{1}{2\pi^2}S_{q=1}'=-\frac{1}{2\pi^2}(-\text{Vol}(\mathbb{H}^{d-1})\frac{\pi^{d/2+1}\Gamma(d/2)(d-1)}{(d+1)!}C_T)|_{d=4}\nn\\&=&-\frac{\pi^2}{40}C_T\log\frac{R}{\epsilon}.
\eea  
In the equation above, we just include the cutoff independent term. It is consistent with \eqref{p1eT1} and \eqref{p1eT2}. Note this is also an independent check for the method of regularization. In the integral \eqref{QAT}, there will be poles when the two stress tensors attach to each other, their effects have been discussed in Appendix \ref{sing}. Since they do not appear in the context of R\'enyi entanglement entropy, it is fine to remove these effects through our regularization. 
\end{enumerate}
\subsubsection{$(3)$-type}
We will consider the following two examples.
\begin{enumerate}
\item Spin 1-1-2. In this case, the three point function is \cite{Osborn:1993cr}
\be
\langle T_{\mu\nu}(x_1)\mathcal{J}_{\sigma}(x_2)\mathcal{J}_{\rho}(x_3)\rangle=\frac{I_{\sigma\alpha}(x_{21})I_{\rho\beta}(x_{31})t_{\mu\nu\alpha\beta}(X_{23})}{x_{12}^d x_{13}^{d}x_{23}^{d-2}},
\ee 
where 
\be
t_{\mu\nu\sigma\rho}(X)=a h^1_{\mu\nu}(\hat{X})\eta_{\sigma\rho}+b h_{\mu\nu}^1(\hat{X})h^1_{\sigma\rho}(\hat{X})+c \ h^2_{\mu\nu\sigma\rho}(\hat{X})+e h^3_{\mu\nu\sigma\rho}
\ee 
with 
\bea
&&h_{\mu\nu}^1(\hat{X})=\hat{X}_{\mu}\hat{X}_{\nu}-\frac{1}{d}\eta_{\mu\nu},\quad \hat{X}_{\mu}=\frac{X_{\mu}}{\sqrt{X^2}},\nn\\
&&h_{\mu\nu\sigma\rho}^2(\hat{X})=\hat{X}_{\mu}\hat{X}_{\sigma}\eta_{\nu\rho}+\hat{X}_{\nu}\hat{X}_{\rho}\eta_{\mu\sigma}+\hat{X}_{\mu}\hat{X}_{\rho}\eta_{\nu\sigma}+\hat{X}_{\nu}\hat{X}_{\sigma}\eta_{\mu\rho}-\frac{4}{d}\hat{X}_{\mu}\hat{X}_{\nu}\eta_{\sigma\rho}-\frac{4}{d}\hat{X}_{\sigma}\hat{X}_{\rho}\eta_{\mu\nu}+\frac{4}{d^2}\eta_{\mu\nu}\eta_{\sigma\rho},\nn\\
&&h_{\mu\nu\sigma\rho}^3=\eta_{\mu\sigma}\eta_{\nu\rho}+\eta_{\mu\rho}\eta_{\nu\sigma}-\frac{2}{d}\eta_{\mu\nu}\eta_{\sigma\rho}. \label{h1h2h3}
\eea
The tensors
$h^1_{\mu\nu}, h^2_{\mu\nu\sigma\rho}, h^3_{\mu\nu\sigma\rho}$ are traceless 
\be
\eta^{\mu\nu}h^1_{\mu\nu}=0,\quad \eta^{\mu\nu}h^2_{\mu\nu\sigma\rho}=0,\quad \eta^{\mu\nu}h^3_{\mu\nu\sigma\rho}=0
\ee
due to the traceless condition of stress tensor. 
The variable 
\be
(X_{23})_{\mu}=\frac{(x_{21})_{\mu}}{x_{21}^2}-\frac{(x_{31})_{\mu}}{x_{31}^2},\quad X_{23}^2=\frac{x_{23}^2}{x_{21}^2x_{31}^2}.
\ee 
The Ward identity from conservation of the currents or stress tensor leads to 
\be
d a-2b+2(d-2)c=0,\quad b-d(d-2)e=0.
\ee 
Only two of the constants are independent. In four dimensions, \be
e=\frac{1}{8}b,\quad c=\frac{1}{2}b-a.
\ee
We only need the component 
\be
\langle T_{00}(x_1)\mathcal{J}_{0}(x_2)\mathcal{J}_{0}(x_3)\rangle=\frac{I_{0\alpha}(x_{21})I_{0\beta}(x_{31})t_{00\alpha\beta}(X_{23})}{x_{12}^d x_{13}^{d}x_{23}^{d-2}}=\frac{t_{0000}(X_{23})}{x_{12}^4 x_{13}^{4}x_{23}^{2}}.
\ee
We notice 
\bea
h_{00}^1(\hat{X})=\frac{1}{4},\quad h_{0000}^2(\hat{X})=\frac{1}{4},\quad h_{0000}^3=\frac{3}{2}.
\eea
Then 
\be
\langle T_{00}(x_1)\mathcal{J}_{0}(x_2)\mathcal{J}_{0}(x_3)\rangle=\frac{-\frac{1}{4}a+\frac{1}{16}b+\frac{1}{4}c+\frac{3}{2}e}{x_{12}^4 x_{13}^{4}x_{23}^{2}}=\frac{3b-4a}{8}\frac{1}{x_{12}^4 x_{13}^{4}x_{23}^{2}}\equiv C_{T\mathcal{J}\mathcal{J}}\frac{1}{x_{12}^4 x_{13}^{4}x_{23}^{2}},
\ee 
where we defined a compact constant $C_{T\mathcal{J}\mathcal{J}}$ which is a linear combination of  $a$ and $b$. Now we can use three different methods to extract the logarithmic term in the CCF of the OPE blocks. 
\begin{enumerate}
\item We can regularize the integral directly 
\bea
&&\langle Q_A[T_{\mu\nu}]Q_A[\mathcal{J}_{\sigma}]^2\rangle_c\nn\\&=&\frac{1}{2}C_{T\mathcal{J}\mathcal{J}}\int_{\Sigma_A} d^3\vec{x}_1 \int_{\Sigma_A}d^3\vec{x}_2 \int_{\Sigma_A}d^3\vec{x}_3 (1-\vec{x}_1^2)\frac{1}{|\vec{x}_1-\vec{x}_2|^4 |\vec{x}_{1}-\vec{x}_3|^{4}|\vec{x}_{2}-\vec{x}_3|^{2}}\nn\\
&=&\frac{1}{2}C_{T\mathcal{J}\mathcal{J}}\int_0^1 r_1^2(1-r_1^2)dr_1 \int_0^1 r_2^2 dr_2\int_0^1 r_3^2 dr_3 \  I_3(2,2,1)
\nn\\&=&8\pi^3 C_{T\mathcal{J}\mathcal{J}}\int_0^{1-\epsilon}dr_1 \frac{r_1^2}{(1-r_1^2)^3}\nn\\&=&\frac{\pi^3}{2}C_{T\mathcal{J}\mathcal{J}}(\frac{R^2}{\epsilon^2}-\frac{R}{\epsilon}-\log\frac{R}{\epsilon}+\cdots).\label{QATJJ}
\eea 
At the second line, we have defined a surface integral $I_3(2,2,1)$ whose details are discussed in the Appendix \ref{int}. Roughly speaking, the integral $I_3(2,2,1)$ has the structure 
\be
I_3(2,2,1)=\tilde{f}_{2,2,1}+\tilde{g}_{2,2,1}\log \frac{r_1+r_2}{|r_1-r_2|}+\tilde{h}_{2,2,1}\log\frac{r_1+r_3}{|r_1-r_3|}+\tilde{i}_{2,2,1}\log\frac{r_2+r_3}{|r_2-r_3|}
\ee 
where the functions $\tilde{f},\tilde{h},\tilde{g},\tilde{i}$ are rational functions of $r_1,r_2,r_3$. Therefore the definite integral becomes elementary. The integrand at the third line has pole $r_1=1$ therefore we insert a small positive UV cutoff $\epsilon$. 
The logarithmic term is indeed has degree 1, 
\be
p_1^e[T_{\mu\nu},\mathcal{J}_{\sigma},\mathcal{J}_{\rho}]=-\frac{\pi^3}{2}C_{T\mathcal{J}\mathcal{J}}.\label{reg1}
\ee 
\item We can compute the following $(2,1)$-type CCF firstly, 
\bea
&&\langle Q_B[T_{\mu\nu}]Q_A[\mathcal{J}_{\sigma}]^2\rangle_c\nn\\&=&\frac{1}{2}C_{T\mathcal{J}\mathcal{J}}\int_{\Sigma_B} d^3\vec{x}_1 \int_{\Sigma_A}d^3\vec{x}_2 \int_{\Sigma_A}d^3\vec{x}_3(1-\vec{x}_1^2)\frac{1}{|\vec{x}_1-\vec{x}_2+\vec{x}_A|^4 |\vec{x}_{1}-\vec{x}_3+\vec{x}_A|^{4}|\vec{x}_{2}-\vec{x}_3|^{2}},\nn\\
\eea
then we can extract the $D$ function by taking the limit $x_A\to\infty$, 
\bea
&&D[\mathcal{J}_{\sigma},\mathcal{J}_{\rho},T_{\mu\nu}]\nn\\&=&\frac{1}{2}C_{T\mathcal{J}\mathcal{J}}\times \frac{1}{2^8}\int_{\Sigma_B} d^3\vec{x}_1 \int_{\Sigma_A}d^3\vec{x}_2 \int_{\Sigma_A}d^3\vec{x}_3(1-\vec{x}_1^2)\frac{1}{|\vec{x}_{2}-\vec{x}_3|^{2}}\nn\\&=&\frac{1}{2^9}C_{T\mathcal{J}\mathcal{J}}\times \frac{2}{15}(4\pi)^2(2\pi)\int_0^1 r_2^2 dr_2  \int_0^1 r_3^2 dr_3\int_0^\pi \sin\theta d\theta \frac{1}{r_2^2+r_3^2-2r_2r_3 \cos\theta}\nn\\&=&\frac{\pi^3}{120}C_{T\mathcal{J}\mathcal{J}}\int_0^1 dr_2 \int_0^1 dr_3 r_2 r_3(\log(r_2+r_3)-\log|r_2-r_3|)\nn\\&=&\frac{\pi^3}{240}C_{T\mathcal{J}\mathcal{J}}.
\eea 
Therefore, using the UV/IR relation we extract the logarithmic term 
\be
p_1^e[\mathcal{J}_{\sigma},\mathcal{J}_{\rho},T_{\mu\nu}]=-120\times\frac{\pi^3}{240}C_{T\mathcal{J}\mathcal{J}}=-\frac{\pi^3}{2}C_{T\mathcal{J}\mathcal{J}}.\label{reg2}
\ee
\item We can also compute another $(2,1)$-type CCF, 
\bea
&&\langle Q_A[T_{\mu\nu}]Q_A[\mathcal{J}_{\sigma}]Q_B[\mathcal{J}_{\rho}]\rangle_c\nn\\&=&\frac{1}{2}C_{T\mathcal{J}\mathcal{J}}\int_{\Sigma_A} d^3\vec{x}_1 \int_{\Sigma_A}d^3\vec{x}_2 \int_{\Sigma_B}d^3\vec{x}_3(1-\vec{x}_1^2)\frac{1}{|\vec{x}_1-\vec{x}_2|^4 |\vec{x}_{1}-\vec{x}_3+\vec{x}_A|^{4}|\vec{x}_{2}-\vec{x}_3+\vec{x}_A|^{2}},\nn\\
\eea 
and read out the large $x_A$ behaviour 
\bea
&&D[T_{\mu\nu},\mathcal{J}_{\sigma},\mathcal{J}_{\rho}]\nn\\&=&\frac{1}{2}C_{T\mathcal{J}\mathcal{J}}\times \frac{1}{2^6}\int_{\Sigma_A} d^3\vec{x}_1 \int_{\Sigma_A}d^3\vec{x}_2 \int_{\Sigma_B}d^3\vec{x}_3(1-\vec{x}_1^2)\frac{1}{|\vec{x}_1-\vec{x}_2|^4}\nn\\&=&\frac{1}{2^7}C_{T\mathcal{J}\mathcal{J}}\times \frac{4\pi}{3} (4\pi)(2\pi) \int_0^1 r_1^2 dr_1 \int_0^1 r_2^2 dr_2 \int_0^\pi \sin\theta d\theta \frac{(1-r_1^2)}{(r_1^2+r_2^2-2r_1r_2\cos\theta)^2}\nn\\&=&-\frac{\pi^3}{24}C_{T\mathcal{J}\mathcal{J}}.
\eea
Now we can extract the logarithmic term 
\be
p_1^e[T_{\mu\nu},\mathcal{J}_{\sigma},\mathcal{J}_{\rho}]=12\times (-\frac{\pi^3}{24}C_{T\mathcal{J}\mathcal{J}})=-\frac{\pi^3}{2}C_{T\mathcal{J}\mathcal{J}}.\label{reg3}
\ee 
\end{enumerate}
Interestingly, the three results \eqref{reg1},\eqref{reg2} and \eqref{reg3} are equal to each other.  This is also the first example that the cyclic identity for $p_1^e$ has been checked.
\item Spin 2-2-2. The three point function of the stress tensor is
\be
\langle T_{\mu\nu}(x_1)T_{\sigma\rho}(x_2)T_{\alpha\beta}(x_3)\rangle=\frac{I_{\mu\nu,\mu'\nu'}(x_{13})I_{\sigma\rho,\sigma'\rho'}(x_{23})t_{\mu'\nu'\sigma'\rho'\alpha\beta}(X_{12})}{ x_{13}^{2d} x_{23}^{2d}},
\ee 
The structure of $t_{\mu\nu\sigma\rho\alpha\beta}(X)$ could be found in \cite{Erdmenger:1996yc}. There are three independent coefficients $\mathcal{A},\mathcal{B},\mathcal{C}$ in the three point function of stress tensor.  In this paper, we just need the component
\bea
\langle T_{00}(x_1)T_{00}(x_2)T_{00}(x_3)\rangle_c=\frac{C_{TTT}}{x_{13}^{d}x^{d}_{23}x^{d}_{12}}
\eea 
with
\be
C_{TTT}=\frac{-2(4-5d+2d^2)\mathcal{A}+d \mathcal{B}+2(5d-4)\mathcal{C}}{4d^2},
\ee 
We can use two different methods to extract the logarithmic term.
\begin{enumerate}
\item The first method is to compute the logarithmic term directly, 
\bea
&&\langle Q_A[T_{\mu\nu}]^3\rangle_c\nn\\&=&\frac{1}{8}C_{TTT}\int_{\Sigma_A}d^3\vec{x}_1 \int_{\Sigma_A}d^3\vec{x}_2 \int_{\Sigma_A}d^3\vec{x}_3 (1-\vec{x}_1^2)(1-\vec{x}_2^2)(1-\vec{x}_3^2)\frac{1}{|\vec{x}_1-\vec{x}_2|^4|\vec{x}_1-\vec{x}_3|^4|\vec{x}_2-\vec{x}_3|^4}\nn\\&=&
\frac{1}{8}C_{TTT}\int_0^1 r_1^2 (1-r_1^2) dr_1\int_0^1 r_2^2 (1-r_2^2)dr_2\int_0^1 r_3^2(1-r_3^2)dr_3 I_3(2,2,2)
\nn\\&=&-\frac{\pi^3}{12}C_{TTT}(\frac{R^2}{\epsilon^2}-\frac{R}{\epsilon}-\log\frac{R}{\epsilon}+\cdots).
\eea 
As previous example, we define the integral $I_3(2,2,2)$ in Appendix \ref{int}.  We also insert a small $\epsilon$ in the integral of $r_1$ at the last step. From the result, we read 
\be
p_1^e[T_{\mu\nu},T_{\rho\sigma},T_{\alpha\beta}]=\frac{\pi^3}{12}C_{TTT}.
\ee 
\item The second method is to use UV/IR relation.  We first read the coefficient $D$ in the large $x_A$ limit, 
\bea
&&D[T_{\mu\nu},T_{\rho\sigma},T_{\alpha\beta}]\nn\\&=&\frac{1}{8\times 2^8}C_{TTT}\int_{\Sigma_A}d^3\vec{x}_1 \int_{\Sigma_A}d^3\vec{x}_2 \int_{\Sigma_A}d^3\vec{x}_3 (1-\vec{x}_1^2)(1-\vec{x}_2^2)(1-\vec{x}_3^2)\frac{1}{|\vec{x}_1-\vec{x}_2|^4}\nn\\&=&\frac{1}{2^{11}}C_{TTT}S_2^2 S_1 \frac{2}{15}\int_0^1 dr_1 (1-r_1^2)r_1^2\int_0^1 dr_2 (1-r_2^2)r_2^2 \int_0^\pi \sin\theta d\theta \frac{1}{(r_1^2+r_2^2-2r_1 r_2 \cos\theta)^2}\nn\\&=&-\frac{\pi^3}{1440}C_{TTT}.
\eea 
Therefore 
\be
p_1^e[T_{\mu\nu},T_{\sigma\rho},T_{\alpha\beta}]=(-120)\times (-\frac{\pi^3}{1440}C_{TTT})=\frac{\pi^3}{12}C_{TTT}.
\ee 
\end{enumerate}
The cutoff independent term is the same for different methods. We also check that the result can be mapped to the second derivative of R\'enyi entanglement entropy \cite{Lee:2014zaa},
\be
 \langle Q_A[T_{\mu\nu}]^3\rangle_c=\langle H_{\tau}^3\rangle_c=\frac{3}{8\pi^3}S_{q=1}''.
\ee 
\end{enumerate}
\subsection{Class II}
A Type-O OPE operator is 
\be
Q_A[\mathcal{O}_{\mu_1\cdots\mu_J}]=\int_{A}d^d x K^{\mu_1}\cdots K^{\mu_J} |K|^{\Delta-J-d} \mathcal{O}_{\mu_1\cdots\mu_J}.
\ee 
We change the coordinates to 
\be
t=\frac{\zeta-\bar{\zeta}}{2},\quad \vec{x}=\frac{\zeta+\bar{\zeta}}{2}\vec{\omega}, \quad \vec{\omega}^2=1.
\ee 
The metric of Minkowski spacetime becomes 
\be
ds^2=d\zeta d\bar{\zeta}+\frac{(\zeta+\bar{\zeta})^2}{4}d\vec{\omega}^2, \quad -1<\zeta,\bar{\zeta}<1.\label{newm}
\ee
The new metric \eqref{newm} covers the diamond $A$ twice, then 
\be
d^dx=(\frac{1}{2})^d |\zeta+\bar{\zeta}|^{d-2} d\zeta d\bar{\zeta}d\vec{\omega}.
\ee 
Then the Type-O OPE becomes 
\bea
&&Q_A[\mathcal{O}_{\mu_1\cdots\mu_J}]\nn\\&=&\frac{1}{2^{\Delta-J}}\int_{\mathbb{D}^2}d\zeta d\bar{\zeta}|\zeta+\bar{\zeta}|^{d-2}(1-\zeta^2)^{\frac{\Delta-J-d}{2}}(1-\bar{\zeta}^2)^{\frac{\Delta-J-d}{2}}\int_{S^{d-2}}d^{d-2}\vec{\omega}K^{\mu_1}\cdots K^{\mu_J}\mathcal{O}_{\mu_1\cdots \mu_J}\nn\\
&=&2^{J-\Delta}\int_{\mathbb{D}^2}d^2\mu_J \int_{S^{d-2}}d\vec{\omega}K^{\mu_1}\cdots K^{\mu_J}\mathcal{O}_{\mu_1\cdots \mu_J}.
\eea
The measure 
\be
d^2\mu_J=d\zeta d\bar{\zeta}|\zeta+\bar{\zeta}|^{d-2}(1-\zeta^2)^{\frac{\Delta-J-d}{2}}(1-\bar{\zeta}^2)^{\frac{\Delta-J-d}{2}}.\label{measure}
\ee
The subscript $J$ is used to label the spin $J$ in the measure. The dimension is understood as $d=4$ in this expression. 
The region $\mathbb{D}^2$ is a square with
\be
-1<\zeta,\bar{\zeta}<1.
\ee
Some integrals used in the following has been discussed in Appendix \ref{int2}. 
\subsubsection{$(2)$-type} 
\begin{enumerate}
\item Spin 0.  
\bea
&&\langle Q_A[\mathcal{O}]^2\rangle_c\nn\\&=&2^{-2\Delta}\int_{\mathbb{D}^2}d^2\mu_0 \int_{\mathbb{D}^2}d^2\mu_0'  \int_{S^2}d\vec{\omega} \int_{S^2}d\vec{\omega}' \frac{N_{\mathcal{O}}}{|x-x'|^{2\Delta}}\nn\\&=&2^{-2\Delta} S_2 S_1 \int_{\mathbb{D}^2}d^2\mu_0\int_{\mathbb{D}^2}d^2\mu_0' \int_0^\pi \sin\theta d\theta \frac{N_{\mathcal{O}}}{(a+b \cos\theta)^{\Delta}}\label{QAO2}
\eea
where we define
\bea
(x-x')^2=a+b\ \omega\cdot\omega'
\eea
with 
\be
a=\zeta\bar{\zeta}+\zeta'\bar{\zeta}'+\frac{1}{2}(\zeta-\bar{\zeta})(\zeta'-\bar{\zeta}'),\quad b=-\frac{1}{2}(\zeta+\bar{\zeta})(\zeta'+\bar{\zeta}').\label{ab}
\ee 
The angular between $\vec{\omega}$ and $\vec{\omega}'$ is denoted as $\theta$. The regularization of \eqref{QAO2} is not easy for general $\Delta$. However, we can compute several examples. 
For $\Delta=4$,
\bea
&&\langle Q_A[\mathcal{O}]^2\rangle_c\nn\\&=&\frac{\pi^2}{48}N_{\mathcal{O}}\int_{\mathbb{D}^2}d\zeta d\bar{\zeta}\int_{\mathbb{D}^2}d\zeta'd\bar{\zeta}'\frac{(3a^2+b^2)(\zeta+\bar{\zeta})^2(\zeta'+\bar{\zeta}')^2}{(a^2-b^2)^3}\nn\\&=&\frac{\pi^2}{6}N_{\mathcal{O}}\int_{-1+\epsilon}^{1-\epsilon}d\zeta'\int_{-1+\epsilon}^{1-\epsilon}d\bar{\zeta}' \frac{(\zeta'+\bar{\zeta}')^2}{(1-\zeta'^2)^2(1-\bar{\zeta}'^2)}\nn\\
&=&\frac{\pi^2}{12}N_{\mathcal{O}}(\frac{R^2}{\epsilon^2}-\frac{R}{\epsilon}-\log^2\frac{R}{\epsilon}+\cdots).
\eea
At the first step, we integrate the angular part. At the second step, we integrate $\zeta,\bar{\zeta}$ part, the integrand becomes singular for 
\be
\zeta=\pm 1 \quad \text{and} \quad \bar{\zeta}=\pm1,
\ee 
therefore we insert a small UV cutoff $\epsilon$ into the integral. Then the final result obeys area law and there is a logarithmic term with degree 2. The $\cdots$ term includes a logarithmic term with power 1 and a constant. Therefore, the cutoff independent information is
\be
p_2^e[\mathcal{O},\mathcal{O}]=-\frac{\pi^2}{12}N_{\mathcal{O}},\quad \Delta=4.\label{p2eO4}
\ee 
The method can be extended to other even conformal weight, for example, 
\be
p_2^e[\mathcal{O},\mathcal{O}]=-\frac{\pi^2}{720}N_{\mathcal{O}},\quad \Delta=6.\label{p2eO6}
\ee
 Now we'd like to use UV/IR relation to obtain this result. 
\bea
&&\langle Q_A[\mathcal{O}]Q_B[\mathcal{O}]\rangle_c\nn\\&=&\int_{A}d^d x \int_A d^d x' |K|^{\Delta-d}|K'|^{\Delta-d}\frac{N_{\mathcal{O}}}{|x-x'|^{2\Delta}}\nn\\&\approx&2^{-2\Delta}S_2^2\int_{\mathbb{D}^2}d^2\mu_0 \int_{\mathbb{D}^2}d^2\mu_0' \frac{N_{\mathcal{O}}}{x_A^{2\Delta}}\nn\\&=&2^{-2\Delta}\times 16\pi^2 \times \frac{N_{\mathcal{O}}}{2^{2\Delta}}z^{\Delta} (H_0)^2\nn\\
&=&D[\mathcal{O},\mathcal{O}]z^\Delta,
\eea 
where 
\be
D[\mathcal{O},\mathcal{O}]=\frac{\pi^2\Gamma(\frac{\Delta}{2}-1)^4\Gamma(\frac{\Delta}{2})^4}{4\Gamma(\Delta)^2\Gamma(\Delta-1)^2}N_{\mathcal{O}}.
\ee 
Note at the second step, we use the approximation that A and B are far apart and only extract the leading order behaviour. At the third step, $H_0$ is defined in Appendix \ref{int2}. Therefore we can use UV/IR relation 
\be
p_2^e[\mathcal{O},\mathcal{O}]=E[\mathcal{O}]D[\mathcal{O},\mathcal{O}]=-\frac{4\pi^2(\Delta-1)\Gamma(\Delta-2)^2\Gamma(\frac{\Delta}{2})^4}{\Gamma(\Delta)^2\Gamma(\Delta-1)^2}N_{\mathcal{O}}.\label{p2eOO}
\ee 
The coefficient \eqref{p2eOO} matches with \eqref{p2eO4} and \eqref{p2eO6} for $\Delta=4$ and 6, correspondingly. Interestingly, we obtain the general coefficient $p_2^e[\mathcal{O},\mathcal{O}]$ from UV/IR relation. This result is not easy to find if we regularize \eqref{QAO2} directly. There are two special points for the coefficient \eqref{p2eOO}. 
\begin{enumerate}
\item $\Delta=1$. In this case, the conformal weight satisfies the unitary bound for scalar operator, $p_2^e[\mathcal{O},\mathcal{O}]=0$. One may need to study the coefficient $p_1^e[\mathcal{O},\mathcal{O}]$ to find the cutoff independent information.
\item $\Delta=2$. In this case, the coefficient $p_2^e[\mathcal{O},\mathcal{O}]\to\infty$. We don't find a way to understand this phenomenon.
\end{enumerate}
\item Spin 1. The CCF is 
\bea
&&\langle Q_A[\mathcal{O}_{\mu}]Q_B[\mathcal{O}_\nu]\rangle_c\nn\\&=&2^{2-2\Delta}N_{\mathcal{O}_{\mu}}\int_{\mathbb{D}^2}d^2\mu_1 \int_{\mathbb{D}^2}d^2\mu'_1\int_{S^2}d\vec{\omega} \int_{S^2}d\vec{\omega}'\frac{K^{\mu}(x)I_{\mu\nu}(x-x')K^{\nu}(x')}{|x-x'|^{2\Delta}},\label{QABOmu}
\eea
where 
\bea
K^{\mu}(x)I_{\mu\nu}(x-x')K^{\nu}(x')=K\cdot K'-\frac{2}{(x-x')^2} K\cdot N\  K'\cdot N.
\eea
We parameterize 
\bea
(x-x')^2&=&a+b\ \vec{\omega}\cdot\vec{\omega}'+e\ \vec{\omega}\cdot \vec{x}_A+f\ \vec{\omega}'\cdot\vec{x}_A+\vec{x}_A^2,\nn\\
K\cdot K'&=&a_1+b_1\ \vec{\omega}\cdot\vec{\omega}',\nn\\
K\cdot N&=&a_2+b_2\ \vec{\omega}\cdot \vec{\omega}'+e_2\ \vec{\omega}\cdot \vec{x}_A,\nn\\
K'\cdot N&=&a_3+b_3\ \vec{\omega}\cdot \vec{\omega}'+e_3\ \vec{\omega}'\cdot \vec{x}_A,
\eea
where the coefficients are
\bea
&&e=\zeta+\bar{\zeta},\quad f=-(\zeta'+\bar{\zeta}'),\nn\\
&&a_1=-\frac{1}{4}(1-\frac{1}{2}(\zeta^2+\bar{\zeta}^2))(1-\frac{1}{2}(\zeta'^2+\bar{\zeta}'^2)),\quad b_1=\frac{1}{16}(\zeta^2-\bar{\zeta}^2)(\zeta'^2-\bar{\zeta}'^2),\nn\\
&&a_2=-\frac{1}{4}(1-\frac{1}{2}(\zeta^2+\bar{\zeta}^2))(\zeta-\bar{\zeta}-\zeta'+\bar{\zeta}')-\frac{1}{8}(\zeta-\bar{\zeta})(\zeta+\bar{\zeta})^2,\nn\\
&&a_3=-\frac{1}{4}(1-\frac{1}{2}(\zeta'^2+\bar{\zeta}'^2))(\zeta-\bar{\zeta}-\zeta'+\bar{\zeta}')+\frac{1}{8}(\zeta'-\bar{\zeta}')(\zeta'+\bar{\zeta}')^2,\nn\\
&&b_2=\frac{1}{8}(\zeta^2-\bar{\zeta}^2)(\zeta'+\bar{\zeta}'),\quad b_3=-\frac{1}{8}(\zeta'^2-\bar{\zeta}'^2)(\zeta+\bar{\zeta}),\nn\\
&& e_2=-\frac{1}{4}(\zeta^2-\bar{\zeta}^2),\quad e_3=-\frac{1}{4}(\zeta'^2-\bar{\zeta}'^2).
\eea 
The coefficients $a, b$ can be found in \eqref{ab}.
When $A$ and $B$ are far away to each other, the leading term is
\bea
&&\langle Q_A[\mathcal{O}_{\mu}]Q_B[\mathcal{O}_\nu]\rangle_c\nn\\&\approx&2^{2-2\Delta}N_{\mathcal{O}_{\mu}}\int_{\mathbb{D}^2}d^2\mu_1 \int_{\mathbb{D}^2}d^2\mu'_1\int_{S^2}d\vec{\omega} \int_{S^2}d\vec{\omega}' \frac{(a_1+b_1 \vec{\omega}\cdot \vec{\omega}')-2 e_2 e_3\ \vec{\omega}\cdot \hat{\vec{x}}_A \vec{\omega}'\cdot \hat{\vec{x}}_A}{2^{2\Delta}}z^\Delta\nn\\&=&
2^{2-4\Delta}S_2^2N_{\mathcal{O}_{\mu}}\int_{\mathbb{D}^2}d^2\mu_1 \int_{\mathbb{D}^2}d^2\mu'_1 \ a_1 z^\Delta.
\eea
Note at the second step, we define the unit vector in the direction of $\vec{x}_A$ as $\hat{\vec{x}}_A$. After some efforts, we find 
\be
D[\mathcal{O}_{\mu},\mathcal{O}_{\nu}]=-\frac{2^{3-4\Delta}\pi^4 \Delta \Gamma(\frac{\Delta-3}{2})^2\Gamma(\frac{\Delta+1}{2})^2}{\Gamma(\frac{\Delta}{2})\Gamma(1+\frac{\Delta}{2})^3}N_{\mathcal{O}_{\mu}},\quad \Delta>3.
\ee 
Therefore 
\be
p_2^e[\mathcal{O}_{\mu},\mathcal{O}_{\nu}]=E[\mathcal{O}_{\mu}]D[\mathcal{O}_{\mu},\mathcal{O}_{\nu}]=-\frac{4^{1-\Delta}\pi^3\Delta\Gamma(\frac{\Delta-3}{2})\Gamma(\frac{\delta+1}{2})}{\Gamma(\frac{\Delta}{2}+1)^2}N_{\mathcal{O}_{\mu}},\quad \Delta>3.
\label{p2eOOmu}
\ee 
We could check this formula \eqref{p2eOOmu} by computing 
\be
\langle Q_A[\mathcal{O}_{\mu}]^2\rangle_c
\ee 
for special values of $\Delta$.  For example, 
\be
\langle Q_A[\mathcal{O}_{\mu}]^2\rangle_c=\frac{\pi^2}{90}N_{\mathcal{O}_{\mu}}(\frac{R^2}{\epsilon^2}-\frac{R}{\epsilon}-\log^2\frac{R}{\epsilon}+\cdots),\quad \Delta=5.
\ee 
The cutoff independent term matches with the general formulae \eqref{p2eOOmu}.
\item Spin 2. Like previous example, we find 
\bea
\langle Q_A[\mathcal{O}_{\mu\nu}]Q_B[\mathcal{O}_{\rho\sigma}]\rangle_c=2^{4-2\Delta}N_{\mathcal{O}_{\mu\nu}}\int_{\mathbb{D}^2}d^2\mu_2\int_{\mathbb{D}^2}d^2\mu'_2\int_{S^2}d\vec{\omega} \int_{S^2}d\vec{\omega}'\frac{(K^{\mu}I_{\mu\rho}K'^{\rho})^2-\frac{1}{4}K^2 K'^2}{|x-x'|^{2\Delta}}.
\eea
From the leading behaviour when $A$ and $B$ are far away, 
\bea
&&D[\mathcal{O}_{\mu\nu},\mathcal{O}_{\rho\sigma}]\nn\\&=&2^{4-4\Delta}N_{\mathcal{O}_{\mu\nu}}\int_{\mathbb{D}^2}d^2\mu_2\int_{\mathbb{D}^2}d^2\mu'_2\int_{S^2}d\vec{\omega} \int_{S^2}d\vec{\omega}'[(a_1+b_1 \vec{\omega}\cdot\vec{\omega}'-2 e_2 e_3\ \vec{\omega}\cdot \hat{\vec{x}}_A \vec{\omega}'\cdot \hat{\vec{x}}_A)^2-\frac{1}{4}K^2 K'^2]\nn\\&=&\frac{3\pi^6 4^{-2\Delta}\Delta^2\text{csc}^2\frac{\pi\Delta}{2} \Gamma(\frac{\Delta}{2}-1)^2}{\Gamma(3-\frac{\Delta}{2})^2\Gamma(\frac{\Delta-3}{2})^2\Gamma(\frac{\Delta+3}{2})^2}N_{\mathcal{O}_{\mu\nu}},\quad \Delta>4.
\eea
The cutoff independent term is
\be
p_2^e[\mathcal{O}_{\mu\nu},\mathcal{O}_{\rho\sigma}]=E[\mathcal{O}_{\mu\nu}]D[\mathcal{O}_{\mu\nu},\mathcal{O}_{\rho\sigma}]=
-\frac{3\pi^2(\Delta-2)\Delta^2\Gamma(\frac{\Delta}{2}-2)^2\Gamma(\frac{\Delta}{2}-1)^2}{64\Gamma(\Delta-4)\Gamma(\Delta+2)}N_{\mathcal{O}_{\mu\nu}},\quad \Delta>4.
\ee 
The formula could be checked for special values of $\Delta$. For example, 
\be
p_2^e[\mathcal{O}_{\mu\nu},\mathcal{O}_{\rho\sigma}]=-\frac{3\pi^2}{2240}N_{\mathcal{O}_{\mu\nu}}.
\ee 
\end{enumerate}
\subsubsection{$(3)$-type}
We consider the following CCF 
\be
\langle Q_A[\mathcal{O}_1]Q_A[\mathcal{O}_2]Q_A[\mathcal{O}_3]\rangle_c=C_{123}\int_A d^4 x_1 \int_A d^4x_2\int_A d^4 x_3 \frac{|K(x_1)|^{\Delta_1-4}|K(x_2)|^{\Delta_2-4}|K(x_3)|^{\Delta_3-4}}{x_{12}^{\Delta_{12,3}}x_{23}^{\Delta_{23,1}}x_{13}^{\Delta_{13,2}}},\label{QA3}
\ee 
where $\Delta_{ij,k}=\Delta_{i}+\Delta_{j}-\Delta_k$. We will use UV/IR relation to find the correlators. Assuming $A$ and $B$ are far away, then 
\bea
&&\langle Q_A[\mathcal{O}_1]Q_A[\mathcal{O}_2]Q_B[\mathcal{O}_3]\rangle_c\nn\\&\approx&C_{123}\int_A d^4 x_1 \int_A d^4 x_2 \int_B d^4 x_3 \frac{|K(x_1)|^{\Delta_1-4}|K(x_2)|^{\Delta_2-4}|K(x_3)|^{\Delta_3-4}}{x_{12}^{\Delta_{12,3} }x_A^{2\Delta_3}}\nn\\&=&\frac{C_{123}\Gamma(\frac{\Delta_3}{2}-1)^2\Gamma(\frac{\Delta_3}{2})^2}{2^{\Delta_1+\Delta_2+\Delta_3+3}\Gamma(\Delta_3)\Gamma(\Delta_3-1)}S_2^2S_1\int_{\mathbb{D}^2}d^2\mu_0\int_{\mathbb{D}^2}d^2\mu_0' \int_0^\pi d\theta\frac{\sin\theta}{(a+b\cos\theta)^{\frac{\Delta_{12,3}}{2}}}z^{\Delta_3}.
\eea
Therefore 
\bea
D[\mathcal{O}_1,\mathcal{O}_2,\mathcal{O}_3]&=&\frac{4C_{123}\pi^3\Gamma(\frac{\Delta_3}{2}-1)^2\Gamma(\frac{\Delta_3}{2})^2}{2^{\Delta_1+\Delta_2+\Delta_3}\Gamma(\Delta_3)\Gamma(\Delta_3-1)}\int_{\mathbb{D}^2}d\zeta d\bar{\zeta}(\zeta+\bar{\zeta})^2\int_{\mathbb{D}^2}d\zeta'd\bar{\zeta}'(\zeta'+\bar{\zeta}')^2\nn\\&&\times (1-\zeta^2)^{\frac{\Delta_1-4}{2}}(1-\bar{\zeta}^2)^{\frac{\Delta_1-4}{2}}(1-\zeta'^2)^{\frac{\Delta_2-4}{2}}(1-\bar{\zeta}'^2)^{\frac{\Delta_2-4}{2}}\int_0^\pi d\theta\frac{\sin\theta}{(a+b\cos\theta)^{\frac{\Delta_{12,3}}{2}}}.\nn\\
\eea 
A close result is not easy to obtain. However, we could find the result case by case, for example \bea
D[\mathcal{O}_1,\mathcal{O}_2,\mathcal{O}_3]=\left\{\begin{array}{l}-\frac{\pi^3}{384}C_{123},\quad \Delta_1=\Delta_2=\Delta_3=4,\vspace{4pt}\\
\frac{\pi^3}{174182400}C_{123},\quad \Delta_1=4,\Delta_2=6,\Delta_3=8,\vspace{4pt}\\
\frac{\pi^3}{4976640}C_{123},\quad \Delta_1=4,\Delta_2=8,\Delta_3=6,\vspace{4pt}\\
\frac{\pi^3}{82944}C_{123},\quad \Delta_1=6,\Delta_2=8,\Delta_3=4,\vspace{4pt}\\
\cdots
\end{array}\right.
\eea 
Then 
\bea
p_2^e[\mathcal{O}_1,\mathcal{O}_2,\mathcal{O}_3]=\left\{\begin{array}{l}\frac{\pi^3}{8}C_{123},\quad \Delta_1=\Delta_2=\Delta_3=4,\vspace{4pt}\\
-\frac{\pi^3}{1728}C_{123},\quad \Delta_1=4,\Delta_2=6,\Delta_3=8,\vspace{4pt}\\
-\frac{\pi^3}{1728}C_{123},\quad \Delta_1=4,\Delta_2=8,\Delta_3=6,\vspace{4pt}\\
-\frac{\pi^3}{1728}C_{123},\quad \Delta_1=6,\Delta_2=8,\Delta_3=4,\vspace{4pt}\\
\cdots
\end{array}\right.
\eea 
Note the last three coefficients are equal which is a consequence of the consistency condition \eqref{p2eeq}.  
We will close this section with some comments on the coefficient $p_2^e$. \begin{enumerate}\item For general conformal weights, we have
\bea
p_2^e[\mathcal{O}_1,\mathcal{O}_2,\mathcal{O}_3]&=&-2^{4-\Delta_1-\Delta_2-\Delta_3}\pi^3C_{123}\int_{\mathbb{D}^2}d\zeta d\bar{\zeta}(\zeta+\bar{\zeta})^2\int_{\mathbb{D}^2}d\zeta'd\bar{\zeta}'(\zeta'+\bar{\zeta}')^2\nn\\&&\times (1-\zeta^2)^{\frac{\Delta_1-4}{2}}(1-\bar{\zeta}^2)^{\frac{\Delta_1-4}{2}}(1-\zeta'^2)^{\frac{\Delta_2-4}{2}}(1-\bar{\zeta}'^2)^{\frac{\Delta_2-4}{2}}\int_0^\pi d\theta\frac{\sin\theta}{(a+b\cos\theta)^{\frac{\Delta_{12,3}}{2}}},\nn\\\label{cyc}
\eea
We don't find an obvious way to prove the cyclic property \eqref{p2eeq} from this expression. It would be quite interesting to check the cyclic property for \eqref{cyc}.
\item In the special case, $\Delta_1+\Delta_2=\Delta_3$, or equivalently, $\Delta_{12,3}=0$, we observe that 
\bea
p_2^e[\mathcal{O}_1,\mathcal{O}_2,\mathcal{O}_3]&=&-2^{5-\Delta_1-\Delta_2-\Delta_3}\pi^3 C_{123}\int_{\mathbb{D}^2}d\zeta d\bar{\zeta}(\zeta+\bar{\zeta})^2\int_{\mathbb{D}^2}d\zeta'd\bar{\zeta}'(\zeta'+\bar{\zeta}')^2\nn\\&&\times (1-\zeta^2)^{\frac{\Delta_1-4}{2}}(1-\bar{\zeta}^2)^{\frac{\Delta_1-4}{2}}(1-\zeta'^2)^{\frac{\Delta_2-4}{2}}(1-\bar{\zeta}'^2)^{\frac{\Delta_2-4}{2}}\nn\\&=&-\frac{\pi^3}{2}\frac{\Gamma(\frac{\Delta_1}{2}-1)^2\Gamma(\frac{\Delta_1}{2})^2\Gamma(\frac{\Delta_2}{2}-1)^2\Gamma(\frac{\Delta_2}{2})^2}{\Gamma(\Delta_1)\Gamma(\Delta_1-1)\Gamma(\Delta_2)\Gamma(\Delta_2-1)}C_{123}. 
\eea 
\end{enumerate}
\section{Discussion}\label{s5}

In the previous section, we have examined the area law of $(m)$-type CCF when all the OPE blocks are the same type. However, we avoid  the following CCF 
\be
\langle Q_A[\mathcal{O}](\cdots) Q_A[\mathcal{J}]\rangle_c, \label{mixtype}
\ee 
where $Q_A[\mathcal{O}]$ is a type-O OPE block while $Q_A[\mathcal{J}]$ is a type-J OPE block. Let's set spacetime dimension to be even and $m=3$.  According to the method of analytic continuation,  we could move either the type-O OPE block or the type-J OPE block to region B, in the first case, we find 
\be
\langle Q_A[\tilde{\mathcal{O}}]Q_A[\mathcal{J}]Q_B[\mathcal{O}]\rangle_c=D[\tilde{\mathcal{O}},\mathcal{J},\mathcal{O}]G_{\Delta,J}(z),\label{case1}
\ee 
where $\Delta$ is the conformal weight of the primary operator $\mathcal{O}$ and $J$ is its spin. In the second case, we find 
\be
\langle Q_A[\mathcal{O}]Q_A[\tilde{\mathcal{O}}]Q_B[\mathcal{J}]\rangle_c=D[\mathcal{O},\tilde{\mathcal{O}},\mathcal{J}]G_{\Delta',J'}(z),\label{case2}
\ee 
where $\Delta'$ is the conformal weight of the primary conserved current $\mathcal{J}$ and $J'$ is its spin. From analytic continuation of \eqref{case1}, we find a $(3)$-type CCF with degree $q=2$, 
\be
p_2^e[\tilde{\mathcal{O}},\mathcal{J},\mathcal{O}]=E[\mathcal{O}]D[\tilde{\mathcal{O}},\mathcal{J},\mathcal{O}]. \label{p2e1}
\ee 
At the same time, from analytic continuation of \eqref{case2}, we find a $(3)$-type CCF with degree $q=1$, 
\be
p_1^e[\mathcal{O},\tilde{\mathcal{O}},\mathcal{J}]=E[\mathcal{J}]D[\mathcal{O},\tilde{\mathcal{O}},\mathcal{J}].\label{p1e2}
\ee 
However, the cutoff independent structure should be the same while \eqref{p2e1} contrasts with \eqref{p1e2} since they predict rather different logarithmic behaviour. This `incompatibility' is based on our implicit assumption that the $D$ function is finite. Actually, as we will show in the following two examples, the $D$ function in \eqref{p1e2} has the following behaviour, 
\be
D[\mathcal{O},\tilde{\mathcal{O}},\mathcal{J}]=D_{\log}[\mathcal{O},\tilde{\mathcal{O}},\mathcal{J}] \log\frac{R}{\epsilon}+\text{finite terms}.
\ee 
The finite terms are cutoff dependent which have no contribution to the degree $q$ of the CCF.  The logarithmic term increases the degree of the CCF by 1, therefore the CCF \eqref{case2} has the same degree $q=2$ as \eqref{case1}.  Instead of \eqref{p1e2}, there should be a modified UV/IR relation 
\be
p_2^e[\mathcal{O},\tilde{\mathcal{O}},\mathcal{J}]=E[\mathcal{J}]D_{\log}[\mathcal{O},\tilde{\mathcal{O}},\mathcal{J}]. \label{p1e3}
\ee
With this modified relation, the two coefficients \eqref{p2e1} and \eqref{p1e3} should be equal to each other
\be
p_2^e[\tilde{\mathcal{O}},\mathcal{J},\mathcal{O}]=p_2^e[\mathcal{O},\tilde{\mathcal{O}},\mathcal{J}].
\ee 
 In other words, they still obey the cyclic identity.  We will use two explicit examples to check this point. 
\begin{enumerate}\item The first CCF we'd like to discuss is
\be
\langle Q_A[T_{\mu\nu}]Q_A[\mathcal{O}]Q_A[\mathcal{O}]\rangle_c,
\ee 
where $T_{\mu\nu}$ is the stress tensor and $\mathcal{O}$ is a spinless primary operator.
The three point function \cite{Osborn:1993cr}
\be
\langle T_{\mu\nu}(x_1)\mathcal{O}(x_2)\mathcal{O}(x_3)\rangle=a\frac{h^1_{\mu\nu}(\hat{X}_{23})}{x_{12}^d x_{23}^{2\Delta-d}x_{13}^d}
\ee 
is fixed up to a theory dependent coefficient $a$.
We just need the $00$ component, it is easy to find 
\be
(\hat{X}_{23})_0(\hat{X}_{23})_0=\frac{x_{21}^2x_{31}^2}{x_{23}^2}\left(\frac{(x_{21})_0}{x_{21}^2}-\frac{(x_{31})_0}{x_{31}^2}\right)^2.
\ee 
Therefore 
\bea
&&\langle Q_A[T_{\mu\nu}]Q_A[\mathcal{O}]Q_B[\mathcal{O}]\rangle_c\nn\\&\approx&\frac{az^\Delta}{2^{1+2\Delta}}\int_{\Sigma_A}d^3\vec{x}_1 (1-\vec{x}_1^2)\int_A d^4 x_2 |K(x_2)|^{\frac{\Delta-4}{2}}\int_B d^4 x_3 |K(x_3)|^{\frac{\Delta-4}{2}} \frac{\frac{t_2^2}{-t_2^2+(\vec{x}_2-\vec{x}_1)^2}+\frac{1}{4}}{(-t_2^2+(\vec{x}_2-\vec{x}_1)^2)^2}\nn\\&=&\frac{\pi 2^{-\Delta-2}\Gamma(\frac{\Delta}{2}-1)^2\Gamma(\frac{\Delta}{2})^2}{\Gamma(\Delta)\Gamma(\Delta-1)}a z^\Delta\int_{\Sigma_A} d^3\vec{x}_1 (1-\vec{x}_1^2) \int_A d^4 x_2 |K(x_2)|^{\frac{\Delta-4}{2}}\frac{\frac{t_2^2}{-t_2^2+(\vec{x}_2-\vec{x}_1)^2}+\frac{1}{4}}{(-t_2^2+(\vec{x}_2-\vec{x}_1)^2)^2}\nn\\&=&\frac{\pi^3 2^{-2\Delta+1}\Gamma(\frac{\Delta}{2}-1)^2\Gamma(\frac{\Delta}{2})^2}{\Gamma(\Delta)\Gamma(\Delta-1)}a z^\Delta\int_0^1 dr_1 r_1^2(1-r_1^2)\int_{\mathbb{D}^2}d^2\mu_0 \int_0^\pi d\theta\frac{\sin\theta}{(r_1^2+\zeta\bar{\zeta}-r_1(\zeta+\bar{\zeta})\cos\theta)^2}\nn\\&&\times[\frac{1}{4}+\frac{(\zeta-\bar{\zeta})^2}{4(r_1^2+\zeta\bar{\zeta}-r_1(\zeta+\bar{\zeta})\cos\theta)}]\nn\\&=&\frac{\pi^3 2^{-2\Delta-1}\Gamma(\frac{\Delta}{2}-1)^2\Gamma(\frac{\Delta}{2})^2}{\Gamma(\Delta)\Gamma(\Delta-1)}a z^\Delta\int_{-1}^1 dr_1 r_1^2(1-r_1^2)\int_{\mathbb{D}^2}d^2\mu_0 \frac{r_1^4-2 \zeta\bar{\zeta}r_1^2+\zeta\bar{\zeta}(\zeta-\bar{\zeta})^2}{(r_1+\zeta)^2(r_1-\zeta)^2(r_1+\bar{\zeta})^2(r_1-\bar{\zeta})^2}\nn\\&=&-\frac{\pi^5 4^{3-2\Delta}\Gamma(\frac{\Delta}{2}-1)^4}{\Delta(\Delta-2)\Gamma(\frac{\Delta-3}{2})\Gamma(\frac{\Delta-1}{2})^2\Gamma(\frac{\Delta+1}{2})}a z^\Delta,\quad \Delta>2.
\eea 
The region $\Delta>2$ is from the convergence of the integral. 
We obtain 
\be
p_2^e[T_{\mu\nu},\mathcal{O},\mathcal{O}]=\frac{2^{5-2\Delta}\pi^4 \Gamma(\frac{\Delta}{2}-1)^2}{\Delta(\Delta-2)\Gamma(\frac{\Delta-3}{2})\Gamma(\frac{\Delta-1}{2})} a,\quad \Delta>2.\label{TOOp2e}
\ee 
As we discussed, we can also compute another CCF 
\bea
&&\langle Q_A[\mathcal{O}]^2Q_B[T_{\mu\nu}]\rangle_c\nn\\&\approx&\frac{az^4}{2^{9}}\int_{\Sigma_B}d^3\vec{x}_1 (1-\vec{x}_1^2)\int_A d^4 x_2 |K(x_2)|^{\frac{\Delta-4}{2}}\int_A d^4 x_3 |K(x_3)|^{\frac{\Delta-4}{2}}\frac{\frac{t_{23}^2}{x_{23}^2}+\frac{1}{4}}{x_{23}^{2\Delta-4}}\nn\\&\equiv&D[\mathcal{O},\mathcal{O},T_{\mu\nu}]z^4.
\eea 
We have defined
\bea
D[\mathcal{O},\mathcal{O},T_{\mu\nu}]&=&\frac{\pi^3 a }{2^{3+2\Delta}\times 15}\int_{\mathbb{D}^2}d^2\mu_0 \int_{\mathbb{D}^2}d^2\mu'_0\int_0^{\pi} d\theta\frac{\sin\theta}{(a+b \cos\theta)^{\Delta-2}}[\frac{1}{4}+\frac{(\zeta-\bar{\zeta}-\zeta'+\bar{\zeta}')^2}{4(a+b\cos\theta)}].\nn\\
\eea 
The integral is not easy, therefore we just compute several examples. Let's set
 $\Delta=4$. Interestingly, we find a logarithmic divergent coefficient $D$, 
\bea
D[\mathcal{O},\mathcal{O},T_{\mu\nu}]&=&-\frac{\pi^3}{3840}a(\log\frac{R}{\epsilon}+\cdots)\label{delta4}
\eea 
where $\cdots$ is a cutoff dependent constant. We can read the coefficient
\be
D_{\log}[\mathcal{O},\mathcal{O},T_{\mu\nu}]=-\frac{\pi^3}{3840}a.
\ee 

Now if we take the limit $B\to A$, the conformal block of the stress tensor will contribute one logarithmic divergence as usual. However, since the coefficient $D$ also has a logarithmic divergence with degree one, there will be a logarithmic term with degree 2 in the final result. Using the modified UV/IR relation,  we get 
\be
p_2^e[\mathcal{O},\mathcal{O},T_{\mu\nu}]=E[T_{\mu\nu}]D_{\log}[\mathcal{O},\mathcal{O},T_{\mu\nu}]=\frac{\pi^3}{32}a.
\ee 
 This is consistent with \eqref{TOOp2e} for $\Delta=4$.  
We could check the logarithmic divergence behaviour for other conformal weights, for example, 
\bea
D[\mathcal{O},\mathcal{O},T_{\mu\nu}]=\left\{\begin{array}{l}-\frac{\pi^3a}{138240}\log\frac{R}{\epsilon}+\cdots,\quad \Delta=6,\vspace{4pt}\\
-\frac{\pi^3a}{4147200}\log\frac{R}{\epsilon}+\cdots,\quad \Delta=8,\vspace{4pt}\\
\cdots
\end{array}\right.
\eea 
Therefore 
\bea
p_2^e[\mathcal{O},\mathcal{O},T_{\mu\nu}]=\left\{\begin{array}{l}\frac{\pi^3}{1152}a ,\quad \Delta=6,\vspace{4pt}\\
\frac{\pi^3}{34560}a,\quad \Delta=8,\vspace{4pt}\\
\cdots
\end{array}\right.\label{p2eTOOre}
\eea 
All the results \eqref{p2eTOOre} are consistent with \eqref{TOOp2e}. 
\item The second CCF we'd like to discuss is 
\be
\langle Q_A[T_{\mu\nu}]Q_A[T_{\sigma\rho}]Q_A[\mathcal{O}]\rangle_c,
\ee 
where $T_{\mu\nu},T_{\sigma\rho}$ are stress tensor and $\mathcal{O}$ is a spin 0 primary operator with conformal weight $\Delta$. The three point function is \cite{Osborn:1993cr}
\be
\langle T_{\mu\nu}(x_1)T_{\sigma\rho}(x_2)\mathcal{O}(x_3)\rangle_c=\frac{I_{\mu\nu,\alpha\beta}(x_{13})I_{\sigma\rho,\gamma\delta}(x_{23})t_{\alpha\beta\gamma\delta}(X_{12})}{x_{12}^{2d-\Delta}x_{23}^{\Delta}x_{13}^{\Delta}},
\ee 
with 
\be
t_{\alpha\beta\gamma\delta}(X)=a h^1_{\alpha\beta}(\hat{X})h^1_{\gamma\delta}(\hat{X})+b \ h^2_{\alpha\beta\gamma\delta}(\hat{X})+c \ h^3_{\alpha\beta\gamma\delta}.
\ee 
The tensors $h_{\mu\nu}^1, h_{\mu\nu\sigma\rho}^2,h_{\mu\nu\sigma\rho}^3$ can be found in \eqref{h1h2h3}.
The conservation of stress tensor leads to two linear relations between $a,b,c$
\bea
&&a+4b-\frac{1}{2}(d-\Delta)(d-1)(a+4b)-d\Delta b=0,\nn\\
&&a+4b+d(d-\Delta)b+d(2d-\Delta)c=0.
\eea
There is an overall constant for the three point function $\langle TT\mathcal{O}\rangle$.  In four dimensions, we find 
\be
b=\frac{10-3\Delta}{4(\Delta-10)}a,\quad c=\frac{3\Delta^2-24\Delta+40}{4(\Delta-8)(\Delta-10)}a.
\ee  
We need the component 
\be
\langle T_{00}(x_1)T_{00}(x_2)\mathcal{O}(x_3)\rangle_c=\frac{\varphi(y)}{x_{12}^{2d-\Delta}x_{23}^\Delta x_{13}^\Delta},\label{TTO3pt}
\ee 
where the function $\varphi$ is 
\be
\varphi(y)=a(\frac{1}{d}+y)^2+4b[\frac{1}{d^2}+\frac{2}{d}y+y(1+2y)]+2c[(1+2y)^2-\frac{1}{d}]
\ee 
with 
\be
y=\frac{t_3^2 x_{12}^2}{x_{13}^2x_{23}^2}.
\ee
Note the time component of $x_1$ and $x_2$ is 0 for \eqref{TTO3pt}. Therefore 
\bea
&&\langle Q_A[T_{\mu\nu}]Q_A[T_{\sigma\rho}]Q_B[\mathcal{O}]\rangle_c\nn\\&\approx&\frac{z^{\Delta}}{2^{3\Delta+2}}\int d^3\vec{x}_1(1-\vec{x}_1^2)\int d^3\vec{x}_2 (1-\vec{x}_2^2)\int_{\mathbb{D}^2}d^2\mu_0 \int_{S^2}d\vec{\omega}_3 \frac{\varphi(y=0)}{x_{12}^{4-\Delta}}.
\eea
We could read 
\be
D[T_{\mu\nu},T_{\sigma\rho},\mathcal{O}]=\frac{1}{2^{3\Delta+2}}H_0 S_2^2 S_1 \int_0^1 dr_1 r_1^2(1-r_1^2) \int_0^1 dr_2 r_2^2(1-r_2^2) \int_0^\pi \sin\theta d\theta \frac{\varphi(y=0)}{(r_1^2+r_2^2-2 r_1 r_2 \cos\theta)^{4-\frac{\Delta}{2}}}.
\ee 
The integral is finite for general $\Delta>3$. 
After some efforts, we find 
\bea
p_2^e[T_{\mu\nu},T_{\sigma\rho},\mathcal{O}]=E[\mathcal{O}]D[T_{\mu\nu},T_{\sigma\rho},\mathcal{O}]=-\frac{4\pi^3 a}{(\Delta+2)\Delta(\Delta-2)(\Delta-8)(\Delta-10)}.\label{p2eTTO}
\eea 
We can also compute the following CCF, 
\bea
&&\langle Q_A[T_{\sigma\rho}]Q_A[\mathcal{O}]Q_B[T_{\mu\nu}]\rangle_c\nn\\&\approx&\frac{z^4}{2^{\Delta+10}}\int d^3\vec{x}_1 \int d^3\vec{x}_2 \int_{\mathbb{D}^2} d^2\mu_0 \int_{S^2}d\vec{\omega}_3 (1-\vec{x}_1^2)(1-\vec{x}_2^2)\frac{\varphi(y)}{x_{23}^\Delta}.
\eea 
The $D$ constant is 
\bea
&&D[T_{\sigma\rho},\mathcal{O},T_{\mu\nu}]\nn\\&=&\frac{1}{2^{\Delta+9}\times 15} S_2^2 S_1\int_0^1 dr_2 r_2^2(1-r_2^2) \int_{\mathbb{D}^2}d^2\mu_0\int_0^\pi \sin\theta d\theta \frac{\varphi(y)}{(r_2^2+\zeta\bar{\zeta}-r_2(\zeta+\bar{\zeta})\cos\theta)^{\frac{\Delta}{2}}},\nn\\\label{TTO3ptT}
\eea
where the function 
\be
y=\frac{(\zeta-\bar{\zeta})^2}{4(r_2^2+\zeta\bar{\zeta}-r_2(\zeta+\bar{\zeta})\cos\theta)}.
\ee 
The integral \eqref{TTO3ptT} is not easy, we choose $\Delta=4$, then 
\be
D[T_{\sigma\rho},\mathcal{O},T_{\mu\nu}]=\frac{\pi^3 a}{34560}\log\frac{R}{\epsilon}+\cdots.
\ee 
It is divergent, therefore the cutoff independent coefficient is $p_2^e$, 
\be
p_2^e[T_{\sigma\rho},\mathcal{O},T_{\mu\nu}]=E[T_{\mu\nu}]D_{\log}[T_{\sigma\rho},\mathcal{O},T_{\mu\nu}]=-\frac{\pi^3 a}{288}.
\ee 
The result is consistent with \eqref{p2eTTO} for $\Delta=4$. We can also calculate other examples, 
\bea
D[T_{\sigma\rho},\mathcal{O},T_{\mu\nu}]=\left\{\begin{array}{l}\frac{\pi^3a}{46800}\log\frac{R}{\epsilon}+\cdots,\quad \Delta=6,\vspace{4pt}\\
\frac{\pi^3a}{403200}\log\frac{R}{\epsilon}+\cdots,\quad \Delta=12,\vspace{4pt}\\
\cdots.\end{array}\right.
\eea 
They are all divergent with a logarithmic term. Then 
\bea
p_2^e[T_{\sigma\rho},\mathcal{O},T_{\mu\nu}]=\left\{\begin{array}{l}-\frac{\pi^3 a}{384},\quad \Delta=6,\vspace{4pt}\\
-\frac{\pi^3a}{3360},\quad \Delta=12,\vspace{4pt}\\
\cdots.\end{array}\right.\label{match}
\eea The result \eqref{match} matches with \eqref{p2eTTO}, correspondingly.
\end{enumerate}

\section{Conclusion}
In this paper, we calculate the divergent behaviour of $(m)$-type CCF of OPE blocks. Due to the complexity of the integrals, we only tackle the case for $m=2$ and $3$. We classify the OPE blocks to type-J and type-O, according to the primary operator in the definition of OPE block.  The logarithmic behaviour has been discussed for varies $(m)$-type CCFs. In even/odd dimensions, we could identify two classes of $(m)$-type CCF according to the degree $q$. 

We establish a formula which is to relate $(m)$-type CCF to $(m-1,1)$-type CCF, we call it UV/IR relation. Schematically, it has the simple form 
\be
p\sim E\times D, \label{ped}
\ee 
where $p$ is the cutoff independent coefficient in $(m)$-type CCF. The coefficient $D$ is  the coefficient before conformal block for $(m-1,1)$-type CCF. The coefficients $p$ and $D$ encode useful information of the CFT. On the other hand, the coefficient $E$ is completely fixed by conformal symmetry, which is a kinematic term. We check the UV/IR relation \eqref{ped} for various examples,  in all cases, the cyclic property of $p$ is always valid, see \eqref{constr} or \eqref{p2eeq}. 

When the OPE blocks belong to different types in $(m)$-type CCF,  the UV/IR relation should be modified to 
\be
p\sim E\times D_{\log}
\ee where $D_{\log}$ is the coefficient before the logarithmic term in the corresponding $D$ function. Note the $(m-1,1)$-type CCF is not always convergent, it may contain logarithmic divergent term in the $D$ coefficient. This is a generalization of the conclusion in \cite{Long:2019pcv} where the author considered $(m-1,1)$-type CCF of type-J OPE blocks in two dimensions. However, we could still obtain cutoff independent coefficient from the logarithmic term in the $D$ coefficient. The cyclic identity is still valid after replacing $D$ by its cutoff independent part $D_{\log}$. 

In all the  examples we compute in this work, we always find $q\le 2$. Since we just consider the cases $m\le 3$, it is not clear whether  $q$ could be larger than 2 or not for general $m$. If  the coefficient $D$ is always finite, then the degree $q$ must be less than or equal to $2$. However, since we find a $(2,1)$-type CCF which shows logarithmic behaviour, it would be quite interesting to explore higher $(m)$-type CCFs.  

Higher $(m)$-type CCF of OPE blocks is also very important to understand the deformed reduced density matrix $\rho_A=e^{-\mu Q_A[\mathcal{O}]}$, a formal exponential non-local operator defined in \cite{Long:2019pcv}. This operator is similar to ``Wilson loop'' \cite{Maldacena:1998im, Rey:1998ik} formally. When the OPE block $Q_A[\mathcal{O}]$ has a lower bound, it is likely that we could read cutoff independent information from the logarithm of the vacuum expectation value of deformed reduced density matrix
\be
\log \langle e^{-\mu Q_A[\mathcal{O}]}\rangle.\label{logexp}
\ee 
A naive continuation from conformal block shows that this quantity \eqref{logexp} also obeys area law \cite{Long:2020njs}. Since conformal block is fixed by conformal invariance, the area law of \eqref{logexp} is protected by conformal symmetry. We'd like to study this point in the future.

\subsection*{Acknowledgements}
This work was supported by NSFC Grant No. 12005069.
\appendix
\section{Singularity}\label{sing}
When two operators attach to each other, there could be singularities. In this Appendix, we will show that these singularities does not affect the cutoff independent coefficient using explicit examples. In \eqref{QAJ}, at the fourth line, the singularities are at $r=r'$, we'd like to examine the singular behaviour carefully. The typical integral is 
\be
I_1=\int_0^1 dr \int_0^1 dr' \frac{r^2 r'^2(r^2+r'^2)}{(r-r')^4(r+r')^4}.
\ee 
We could separate the singularity by replacing the integral by 
\be
I_1=\int_0^1 dr (\int_0^{r-\epsilon}dr'+\int_{r+\epsilon}^1dr')\frac{r^2 r'^2(r^2+r'^2)}{(r-r')^4(r+r')^4}=\int_0^1 dr \frac{r^2}{12\epsilon^3}+I_1'.\label{I1s}
\ee 
The integral $I_1'$ is the one used at the fifth line of \eqref{QAJ}. The first term on the right hand side of \eqref{I1s} is the effect of the singularity, it has been removed from the regularization method in the context. It is easy to find 
\be
I_1=\frac{1}{36\epsilon^3}+I_1',
\ee 
there is no extra logarithmic term. Therefore we conclude that the terms that have been removed do not affect the cutoff indepdendent coefficient. In the same way,  the singularity in \eqref{QAT} is also $r=r'$, the relevant integral is 
\bea
I_2&=&\int_0^1 dr \int_0^1 dr' \frac{r^2(1-r^2)r'^2(1-r'^2)(r^2+3r'^2)(r'^2+3r^2)}{(r^2-r'^2)^6}\nn\\&=&\int_0^1 dr[ \frac{r^2(1-r^2)^2}{10\epsilon^5}+\frac{r^2(1-r^2)}{2\epsilon^3}]+I_2'\nn\\&=&\frac{4}{525\epsilon^5}-\frac{1}{15\epsilon^3}+I_2'.
\eea 
It is obvious that the singularity does not affect the cutoff independent coefficient. 
\section{Integrals}
\subsection{Surface $S^2$}\label{int}
The typical integrals used in this paper is 
\be
I_n(\alpha_{ij})=\prod_{i=1}^n\int_{S^2}d^2\vec{\omega}_i\  \prod_{i<j} |\vec{x}_i-\vec{x}_j|^{-2\alpha_{ij}},\quad n\ge2.
\ee 
where $\vec{x}_i=r_i \vec{\omega}_i$. The integrand only depends on the angle between vectors $\vec{\omega}_i$ and $\vec{\omega}_j$. The constants $\alpha_{ij}$ are assumed to be real. If some of them are positive, then the integral has poles. We assume $r_1>r_2>r_3$ to avoid the pole and this doesn't lose any information of the integral.  For $n=2$, the integral is elementary. In this paper, we need the result for $n=3$. We expand the function $|\vec{x}_i-\vec{x}_j|^{-2\alpha_{ij}}$ in terms of Legendre function of the first kind
\be
|\vec{x}_i-\vec{x}_j|^{-2\alpha_{ij}}=\sum_{\ell=0}^\infty f_{\ell}(r_i,r_j,\alpha_{ij})P_{\ell}(\cos\psi_{ij})
\ee 
with 
\be
\cos\psi_{ij}=\cos\theta_i\cos\theta_j+\sin\theta_i\sin\theta_j\cos(\phi_i-\phi_j).
\ee 
The function $f_{\ell}$ is 
\bea
f_{\ell}(r_i,r_j,\alpha_{ij})&=&\frac{2\ell+1}{2}\int_{-1}^1 dx P_{\ell}(x)(r_i^2+r_j^2-2r_i r_j x)^{-\alpha_{ij}}\nn\\&=&-e^{-i\pi\alpha_{ij}}\frac{(2\ell+1)(z_{ij}^2-1)^{-\frac{\alpha_{ij}-1}{2}}}{(2r_ir_j)^{\alpha_{ij}}\Gamma(\alpha_{ij})}Q_{\ell}^{\alpha_{ij}-1}(z_{ij})\nn\\&=&-e^{-i\pi\alpha_{ij}}\frac{2\ell+1}{\Gamma(\alpha_{ij})}\frac{(r_i^2-r_j^2)^{1-\alpha_{ij}}}{2r_ir_j} Q_{\ell}^{\alpha_{ij}-1}(z_{ij}).
\eea 
At the first line, we used the orthogonal relation of Legendre function of the first kind 
\be
\int_{-1}^1 P_{\ell}(x)P_{\ell'}(x)dx=\frac{2}{2\ell+1}\delta_{\ell,\ell'}.
\ee 
At the second step,  we used the integral formula \cite{tables}
\be
\int_{-1}^1 dx P_{\ell}(x)(z-x)^{-\mu-1}=\frac{2e^{-i\pi\mu}}{\Gamma(1+\mu)}(z^2-1)^{-\frac{\mu}{2}}Q_{\ell}^{\mu}(z).
\ee 
The parameter $z_{ij}=\frac{r_i^2+r_j^2}{2r_ir_j}$.
Since Legendre function of the first kind can be expanded into spherical harmonics
\be
P_{\ell}(\cos\psi_{ij})=\frac{4\pi}{2\ell+1}\sum_{m=-\ell}^{\ell}Y_{\ell m}(\theta_i,\phi_i)Y^*_{\ell m}(\theta_j,\phi_j),
\ee 
Using the orthogonal relation of spherical harmonics 
\be
\int_0^\pi \sin\theta d\theta \int_0^{2\pi} d\phi Y_{\ell m}(\theta,\phi)Y^*_{\ell'm'}(\theta,\phi)=\delta_{\ell,\ell'}\delta_{m,m'},
\ee  
the integral for $n=3$ becomes 
\bea
&&I_3(\alpha_{12},\alpha_{13},\alpha_{23})\nn\\&=& \sum_{\ell=0}^\infty\frac{(4\pi)^3}{(2\ell+1)^2}f_{\ell}(r_1,r_2,\alpha_{12})f_{\ell}(r_1,r_3,\alpha_{13})f_{\ell}(r_2,r_3,\alpha_{23})\nn\\&=&-\frac{8\pi^3 e^{-i\pi(\alpha_{12}+\alpha_{13}+\alpha_{23})}}{\Gamma(\alpha_{12})\Gamma(\alpha_{13})\Gamma(\alpha_{23})}\frac{(r_1^2-r_2^2)^{1-\alpha_{12}}(r_1^2-r_3^2)^{1-\alpha_{13}}(r_2^2-r_3^2)^{1-\alpha_{23}}}{r_1^2r_2^2r_3^2}J(\alpha_{12},\alpha_{13},\alpha_{23}).
\eea 
The problem is reduced to an infinite sum of triple products of associate Legendre Polynomials of the second kind \footnote{In general, the three variables $z_{12},z_{13},z_{23}$ are not related to each other. In our case, they are constrainted by the identity 
	\be z_{12}^2+z_{13}^2+z_{23}^2-1-2 z_{12}z_{13}z_{23}=0.\ee}
\be
J(\alpha_{12},\alpha_{13},\alpha_{23})= \sum_{\ell=0}^{\infty}(2\ell+1)Q_{\ell}^{\alpha_{12}-1}(z_{12})Q_{\ell}^{\alpha_{13}-1}(z_{13})Q_{\ell}^{\alpha_{23}-1}(z_{23}).
\ee 
While the infinite sum of the triple product of Legendre Polynomials of first kind has been found long time ago\cite{Watson:1933},  we don't find a close formula for general $\alpha_{12},\alpha_{13},\alpha_{23}$. Fortunately, we just need the result for special value of $\alpha_{12},\alpha_{13},\alpha_{23}$. With some efforts, the general structure of $J$ is as follows for positive integer $\alpha_{12},\alpha_{13},\alpha_{23}$ 
\be
J(\alpha_{12},\alpha_{13},\alpha_{23})=f_{\alpha_{12},\alpha_{13},\alpha_{23}}+g_{\alpha_{12},\alpha_{13},\alpha_{23}}\log\frac{r_1+r_2}{|r_1-r_2|}+h_{\alpha_{12},\alpha_{13},\alpha_{23}}\log\frac{r_1+r_3}{|r_1-r_3|}+i_{\alpha_{12},\alpha_{13},\alpha_{23}}\log\frac{r_2+r_3}{|r_2-r_3|},
\ee
where $f,g,h,i$ are rational functions of $r_1,r_2,r_3$.
Several examples are
\bea
f_{1,1,1}&=&0,\nn\\
g_{1,1,1}&=&\frac{4r_1r_2r_3^2}{(r_1^2-r_3^2)(r_2^2-r_3^2)},\nn\\
h_{1,1,1}&=&g_{1,1,1}(r_2\leftrightarrow r_3),\nn\\
i_{1,1,1}&=&g_{1,1,1}(r_1\leftrightarrow r_3),\\
f_{2,2,1}&=&\frac{32r_1^2r_2^2r_3^2[(r_1^4+r_2^2r_3^2)(r_2^2+r_3^2)-4r_1^2(r_2^4-r_2^2r_3^2+r_3^4)]}{15(r_1^2-r_2^2)^2(r_1^2-r_3^2)^2(r_2^2-r_3^2)^2},\nn\\
g_{2,2,1}&=&\frac{16r_1(r_1-r_2)r_2(r_1+r_2)r_3^4(5r_1^2r_2^2-r_1^2r_3^2+r_2^2r_3^2-5r_3^4)}{15(r_1^2-r_2^2)^3(r_3^2-r_2^2)^3},\nn\\
h_{2,2,1}&=&g_{2,2,1}(r_2\leftrightarrow r_3),\nn\\
i_{2,2,1}&=&\frac{4r_1^2r_2r_3(15(r_1^8+r_2^4r_3^4)+10r_1^2(r_1^4+r_2^2r_3^2)(r_2^2+r_3^2-r_1^4(r_2^4+68r_2^2r_3^2+r_3^4))}{15(r_1^2-r_2^2)^3(r_1^2-r_3^2)^3},\\
f_{2,2,2}&=&-\frac{8r_1^2r_2^2r_3^2}{105(r_1^2-r_2^2)^3(r_1^2-r_3^2)^3(r_2^2-r_3^2)^3}[113(r_2^8r_3^4+r_2^4r_3^8+r_1^8r_2^4+r_1^4r_2^8+r_1^4r_3^8+r_1^8r_3^4)\nn\\&&-34(r_1^6r_2^6+r_1^6r_3^6+r_2^6r_3^6+r_1^8r_2^2r_3^6+r_1^2r_2^8r_3^2+r_1^2r_2^2r_3^8)-350(r_1^6r_2^4r_3^2+r_1^4r_2^6r_3^2+r_1^6r_2^2r_3^4\nn\\&&+r_1^2r_2^6r_3^4+r_1^4r_2^2r_3^6+r_1^2r_2^4r_3^6)+1626r_1^4r_2^4r_3^4],\nn\\
g_{2,2,2}&=&\frac{16r_1r_2(r_1^2-r_2^2)r_3^4}{105(r_1^2-r_2^2)^4(r_2^2-r_3^2)^4}[35(r_1^4r_2^4+r_3^8)+14(r_1^4r_2^2r_3^2+r_1^2r_2^4r_3^2+r_1^2r_3^6+r_2^2r_3^6)\nn\\&&-r_3^4(r_1^4+r_2^4)-124r_1^2r_2^2r_3^4],\nn\\
h_{2,2,2}&=&g_{2,2,2}(r_2\leftrightarrow r_3),\nn\\
i_{2,2,2}&=&g_{2,2,2}(r_1\leftrightarrow r_3).
\eea  

\subsection{Square $\mathbb{D}^2$}\label{int2}
 The first integral we will use is 
 \be
 H_J=\int_{\mathbb{D}^2}d^2\mu_J.
 \ee
 The measure $d^2\mu_J$ has been defined in \eqref{measure}. The square $\mathbb{D}^2$ is parameterized by two coordintes $\zeta$ and $\bar{\zeta}$
 \be
 -1<\zeta,\bar{\zeta}<1.
 \ee 
 By changing the variables $\zeta,\bar{\zeta}$ to $\xi,\bar{\xi}$ 
 \be
 \xi=\frac{1+\zeta}{2},\quad \bar{\xi}=\frac{1-\bar{\zeta}}{2},
 \ee 
 the integral becomes a standard Selberg integral 
 \bea
 H_J&=&2^{2+d-2+2(\Delta-d-J)}\int_0^1 d\xi \int_0^1 d\bar{\xi}\ |\xi-\bar{\xi}|^{d-2}(\xi(1-\xi)\bar{\xi}(1-\bar{\xi}))^{\frac{\Delta-d-J}{2}}\nn\\&=&2^{2\Delta-d-2J}\text{Sel}_2(1+\frac{\Delta-d-J}{2},1+\frac{\Delta-d-J}{2},\frac{d-2}{2}).
 \eea 
 Selberg integral is defined as \cite{Selberg:1944, For}
 \bea
 \text{Sel}_n(\alpha,\beta,\gamma)&=&\prod_{i=1}^n\int_0^1 d\xi_i \prod_{i=1}^n \xi^{\alpha-1}(1-\xi)^{\beta-1}\prod_{1\le i<j\le n} |\xi_i-\xi_j|^{2\gamma}\nn\\&=&\prod_{j=0}^{n-1}\frac{\Gamma(\alpha+j\gamma)\Gamma(\beta+j\gamma)\Gamma(1+(j+1)\gamma)}{\Gamma(\alpha+\beta+(n+j-1)\gamma)\Gamma(1+\gamma)}.
 \eea
 Therefore 
 \be
 H_J=\frac{2^{2\Delta-d-2J}\Gamma(d-1)\Gamma(\frac{\Delta-J)}{2})^2\Gamma(\frac{\Delta-d-J+2}{2})^2}{\Gamma(\frac{d}{2})\Gamma(\Delta-J)\Gamma(\Delta-J-\frac{d}{2}+1)}.
 \ee 
 In four dimensions, it is 
 \be
 H_J=\frac{2^{2\Delta-2J-3}\Gamma(\frac{\Delta-J-2}{2})^2\Gamma(\frac{\Delta-J}{2})^2}{\Gamma(\Delta-J-1)\Gamma(\Delta)}.
 \ee

\end{document}